\documentclass[12pt,dvipsnames]{article}
\usepackage[top=80pt,bottom=85pt,left=85pt,right=85pt]{geometry}

\pdfoutput=1
\usepackage[utf8]{inputenc}
\usepackage[T1]{fontenc} 
\usepackage[english]{babel} 
\usepackage{braket}
\usepackage[dvipsnames]{xcolor} %Package for fancy coloros
\usepackage{physics}
\usepackage{tabularx} 
\usepackage{ragged2e}
\usepackage{booktabs} 
\usepackage[bb=boondox]{mathalfa}
\usepackage{rotating}
\usepackage{footmisc}
\usepackage{makecell,booktabs}
\usepackage{multirow} 
\usepackage{comment} 
\usepackage{calc}
\usepackage{amsmath}
\usepackage{cite}
\usepackage[all]{xy}
\usepackage{array}

\usepackage{tabularray}

\usepackage{tikz}
\usetikzlibrary{math} 
\usetikzlibrary{3d} 
\usepackage{tikz-3dplot}
\usepackage{tikz-cd} 
\usepackage{microtype}
\usepackage{amsthm}
\usepackage{mathrsfs} 
\usepackage[strict]{changepage}
\newcommand\scalemath[2]{\scalebox{#1}{\mbox{\ensuremath{\displaystyle #2}}}}

\tikzcdset{
  cells={font=\everymath\expandafter{\the\everymath\displaystyle}},
}

\newcommand{\vardbtilde}[1]{\tilde{\raisebox{0pt}[0.85\height]{$\tilde{#1}$}}}

\newcommand{\xdownarrow}[1]{%
  {\left\downarrow\vbox to #1{}\right.\kern-\nulldelimiterspace}
}

\usepackage{amssymb} 
\usepackage{subcaption} 
\DeclareCaptionFormat{custom}

\usepackage[debug,pageanchor=false]{hyperref}
\definecolor{link}{rgb}{.8,.15,.1}
\definecolor{pigment}{rgb}{0.36, 0.54, 0.66}
\definecolor{pigment2}{rgb}{0.19, 0.55, 0.91}
\definecolor{pigment3}{rgb}{0.2, 0.2, 0.6}
\definecolor{light-gray}{gray}{0.75}
\hypersetup{colorlinks=true,linkcolor=link,citecolor=link,urlcolor=link,linktocpage}

\usepackage{float} 
\usepackage{todonotes}

\usepackage[capitalise]{cleveref}

\newcolumntype{E}{>{\hfil$}p{0.65cm}<{$\hfil}}

\newcolumntype{L}{>{\hfil$}p{16cm}<{$\hfil}}

\newcolumntype{D}{>{\hfil$}p{7.4cm}<{$\hfil}}
\newcolumntype{C}{>{\hfil$}p{3cm}<{$\hfil}}
\newcolumntype{P}{>{\hfil$}p{7.7cm}<{$\hfil}}
\newcolumntype{F}{>{\hfil$}p{5.7cm}<{$\hfil}}
\newcolumntype{S}{>{\hfil$}p{1.8cm}<{$\hfil}}
\newcolumntype{R}{>{\hfil$}p{5.2cm}<{$\hfil}}
\newcolumntype{U}{>{\hfil$}p{4.2cm}<{$\hfil}}
\newcolumntype{Q}{>{\hfil$}p{6.4cm}<{$\hfil}}
\newcolumntype{T}{>{\hfil$}p{1.9cm}<{$\hfil}}
\newcolumntype{V}{>{\hfil$}p{5.8cm}<{$\hfil}}
\newcolumntype{H}{>{\hfil$}p{1.8cm}<{$\hfil}}
\newcolumntype{A}{>{\hfil$}p{6cm}<{$\hfil}}
\newcolumntype{B}{>{\hfil$}p{2cm}<{$\hfil}}
\makeatletter
\newcommand\xleftrightarrow[2][]{%
  \ext@arrow 9999{\longleftrightarrowfill@}{#1}{#2}}
\newcommand\longleftrightarrowfill@{%
  \arrowfill@\leftarrow\relbar\rightarrow}
\makeatother

\newcommand\Tstrut{\rule{0pt}{2.6ex}}         % = `top' strut

\numberwithin{equation}{section}

\definecolor{cambridgeblue}{rgb}{0.64, 0.76, 0.68}
\definecolor{caribbeangreen}{rgb}{0.0, 0.8, 0.6}
\definecolor{celadon}{rgb}{0.67, 0.88, 0.69}
\definecolor{champagne}{rgb}{0.97, 0.91, 0.81}
\definecolor{cream}{rgb}{1.0, 0.99, 0.82}
\definecolor{cyan(process)}{rgb}{0.0, 0.72, 0.92}
\definecolor{brilliantlavender}{rgb}{0.96, 0.73, 1.0}
\definecolor{candypink}{rgb}{0.89, 0.44, 0.48}

\begin{document}

\begin{titlepage}

\phantom{wowiezowie}

\vspace{-1cm}

\begin{center}

{\Huge {\bf M-theory geometric engineering for}}

\bigskip

{\Huge {\bf rank-0 3d $\mathcal{N}=2$ theories}}

\vspace{1cm}

{\Large  Andrea Sangiovanni$^{a,b,c}$ and Roberto Valandro$^{d,e}$}\\ 

\medskip

\vspace{1cm}

{\it
{\small

$^a$ Department of Mathematics, Uppsala University,\\ Box 480, SE-75106 Uppsala, Sweden\\
$^b$ Centre for Geometry and Physics, Uppsala University,\\ Box 480, SE-75106 Uppsala, Sweden\\
$^c$ Department of Physics and Astronomy, Uppsala University,\\ Box 516, SE-75120 Uppsala, Sweden\\
%\vspace{.25cm}
$^d$ Dipartimento di Fisica, Universit\`a di Trieste, Strada Costiera 11, I-34151 Trieste, Italy \\%and \\
%\vspace{.25cm}
$^e$ INFN, Sezione di Trieste, Via Valerio 2, I-34127 Trieste, Italy	\\
\vspace{.25cm}

%\vspace{.25cm}
}}

\vskip .5cm
{\footnotesize \tt andrea.sangiovanni@math.uu.se \hspace{1cm} roberto.valandro@ts.infn.it } \\

\vskip 1cm
     	{\bf Abstract }
\vskip .1in

\end{center}

\noindent  M-theory geometric engineering on non-compact Calabi-Yau fourfolds (CY4) produces 3d theories with 4 supercharges. Carefully establishing a dictionary between the geometry of the CY4 and the QFT in the transverse directions remains, to a large extent, an unresolved challenge, complicated by subtleties arising from M5-brane instanton corrections. Such difficulties can be circumvented in the restricted and yet controlled setting offered by CY4 with terminal singularities, as they do not admit crepant resolutions with compact exceptional divisors. After a general review of their properties and partial classifications, we focus on a subclass of terminal CY4 constructed as deformed Du Val singularities, that admit crepant resolutions with at most exceptional 2-cycles. We extract the corresponding 3d $\mathcal{N}=2$ supersymmetric theory descendant in an unambiguous fashion, as the absence of compact 4-cycles leaves no room for a choice of background $G_4$ flux. These turn out to be theories of chiral multiplets with no gauge group and at most abelian flavor factors: we argue that they serve as the simplest building blocks to substantiate a rigorous CY4/3d QFT geometric engineering mapping.

\eject

\end{titlepage}

% titlepage 
\tableofcontents

\section{Introduction}
\indent The string theory geometric engineering philosophy has fostered great advancements in the understanding of the landscape of QFTs in a varying number of dimensions, employing the constraining power of supersymmetry and focusing on conformal theories. Six dimensional (6d) $\mathcal{N}=(2,0)$ superconformal field theories (SCFTs) have been exhaustively classified via Type IIB string theory on backgrounds involving K3 non-compact surfaces with ADE singularities \cite{Witten:1995ex,Strominger_1995}. Similarly, F-theory on elliptically fibered non-compact Calabi-Yau (CY) threefolds has furnished the underpinnings for the geometric realization of 6d $\mathcal{N}=(1,0)$ SCFTs, revealing an elegant classification in terms of fundamental building blocks that can be gauged together, while preserving conformal symmetry \cite{DelZotto:2014hpa,Heckman:2015bfa}.\\
\indent In one dimension less, the recent years have witnessed a sprawling development in the exploration of 5d $\mathcal{N}=1$ theories \cite{Seiberg:1996bd,Morrison:1996xf,Intriligator:1997pq}, employing an interconnected deck of techniques, such as M-theory geometric engineering on non-compact Calabi-Yau threefolds (CY3), $(p,q)$ 5-brane webs, and dimensional reduction from 6d. Despite these efforts, no complete classification is yet in sight. Similarly, theories with 8 supercharges have been vastly studied in 4d (via e.g.\ class $\mathcal{S}$ constructions, D3-branes at singularities and more) and in 3d, unveiling surprising relationships with higher-dimensional SCFTs, as exemplified by the ever-growing literature on magnetic quivers.
\\

\indent Theories with less amount of supercharges are substantially less tractable. The focus of this work lies on \textit{3d theories with 4 supercharges, engineered by M-theory on non-compact CY fourfolds (CY4)}. Much like in the M-theory on CY3 geometric engineering case, it is generally thought that the absence of scales in the singular phase of the CY4 generates a SCFT in the transverse directions.\\
\indent The field-theoretic approach to 3d $\mathcal{N}=2$ theories \cite{Aharony:1997bx} has produced a deluge of results, characterizing dualities between superficially different theories, developing the computation of the superconformal index in a variety of backgrounds, as well as countless other applications.\\
\indent Despite this impressive understanding, and given the low amount of supersymmetry, the analysis of 3d $\mathcal{N}=2$ theories via geometric engineering is a research area that is still in its infancy. The pioneering work of \cite{Gukov:1999ya} considered F-, M- and Type IIA engineering on a specific class of isolated fourfold singularities. More recently, \cite{Najjar:2023hee} have proposed an ambitious program for examining 3d $\mathcal{N}=2$ supersymmetric theories that can be engineered by toric CY fourfolds. Much less, instead, is known for non-toric setups.\\
\indent Operating in this context, throughout this work we posit a two-fold objective:
\begin{itemize}
    \item Systematically organize the mathematical literature on Calabi-Yau fourfolds with (not necessarily isolated) singularities, in view of their physics applications. We devote specific attention to \textit{terminal}\footnote{We will come back to the rigorous definition of terminal singularities momentarily.} singular CY fourfolds, examining both toric and non-toric settings. Terminal singularities do not admit any crepant resolution that produces exceptional divisors, and thus evade the effects produced by M5-brane instantons. Nonetheless, they play a central role for the physics of M-theory on CY4, akin to terminal singularities in CY3. A generic singular CY3 (CY4) can be resolved in a two-step process:
    first, a partial crepant resolution inflates exceptional 4-cycles (6-cycles) corresponding to compact divisors; then, the resulting partially resolved CY3 (CY4) still generally exhibits \textit{terminal} singularities, that might admit small crepant resolutions inflating additional 2-cycles (2-cycles and 4-cycles). See Figure \ref{fig:diagram terminal}.
    The two resolution steps are not always possible, and if they are, they are generally not unique.
    \begin{figure}[H]
    \centering
    \scalebox{0.9}{\begin{tikzpicture}
    \node at (4.5,4.5) {Singular CY3 $X_3$};
    \node at (13.5,4.5) {Singular CY4 $X_4$};
    \draw[->] (4.5,4.2)--(4.5,2.2);
    \draw[->] (13.5,4.2)--(13.5,2.2);
    \node[align=left] at (6.8,3.2) {Partial resolution with\\ exceptional 4-cycles};
    \node[align=left] at (15.8,3.2) {Partial resolution with\\ exceptional 6-cycles};
    \node[align=center] at (4.5,1.5) {Singular CY3 $\tilde{X}_3$\\with \textbf{terminal} singularities};
    \node[align=center] at (13.5,1.5) {Singular CY4 $\tilde{X}_4$\\with \textbf{terminal} singularities};
    \draw[->] (4.5,0.8)--(4.5,-1.2);
    \draw[->] (13.5,0.8)--(13.5,-1.2);
    \node[align=left] at (6.8,-0.2) {Small resolution with\\ exceptional 2-cycles};
    \node[align=left] at (16.1,-0.2) {Small resolution with\\ exceptional 2- and 4-cycles};
    \node[align=center] at (4.5,-1.6) {Smooth CY3 $\vardbtilde{X}_3$};
    \node[align=center] at (13.5,-1.6) {Smooth CY4 $\vardbtilde{X}_4$};
    \end{tikzpicture}}\;
    \caption{Schematic sketch of the crepant resolution procedure for \textit{completely smoothable} CY3 and CY4. For more general singular CY, it might be the case that not every step exists, namely that only a partial resolution or no resolution at all is admitted.}
    \label{fig:diagram terminal}
    \end{figure}
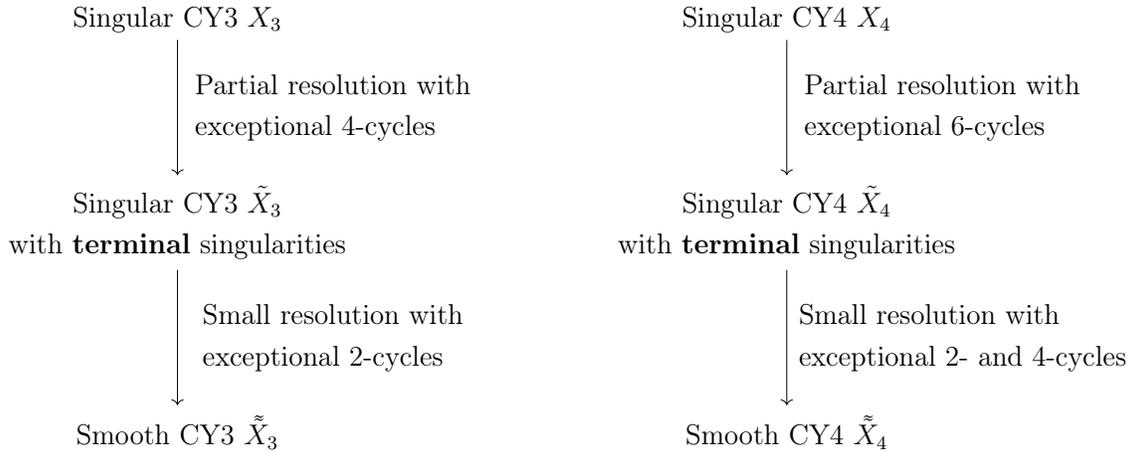    
    \indent It is therefore clear that terminal CY4 singularities appear ubiquitously in the resolution of more general singular CY4, and thus understanding their features is pivotal if any progress towards understanding the M-theory geometric engineering dictionary is to be made. To this end, we review the precise definition of the various beasts in the zoo of terminal Calabi-Yau fourfolds, often drawing comparisons to the singular CY threefold case, highlighting (dis)similarities. We deem this review part as crucial for laying the foundations for a comprehensive study of M-theory geometric engineering on Calabi-Yau fourfolds, both in this and in future works.
    \item We initiate reaping the fruits of the program outlined in the first part by investigating the 3d $\mathcal{N}=2$ theories engineered by a class of singular Calabi-Yau fourfolds, that admit \textit{at most} curves as their exceptional set. These are a generalization of the well-known compound Du Val (cDV) singularities to the fourfold setting. They are also known as ccDV singularities, and they take the form of a Du Val singularity deformed by two-complex parameters:
    \begin{equation}
        \text{ccDV singularities: }\quad P_{\mathfrak{g}}(x,y,z)+h(x,y,z,w,t)=0,
    \end{equation}
    where $P_{\mathfrak{g}}$ is a Du Val singularity labelled by $\mathfrak{g} \in ADE$, and $h$ is a holomorphic function such that $h(x,y,z,0,0) =0$.
    %In addition, there might be some leftover singularity that cannot be crepantly resolved. 
    We will focus on a rich subset of these fourfold singularities, claiming that the 3d theories engineered by M-theory on them satisfy the correct constraints to flow to a SCFT in the IR. They contain uncharged chiral multiplets, or chiral multiplets charged under some abelian flavor group, and \textit{no vector multiplets}. As such, they play the same role, in the 3d $\mathcal{N}=2$ landscape, as their 5d $\mathcal{N}=1$ rank-0 cousins. Namely, given a generic singular Calabi-Yau fourfold admitting a partial crepant resolution producing compact divisors, it is expected that the rank-0 singularities we study in this work generally appear as leftover singularities.
\end{itemize}
We will introduce the appropriate terminology for dealing with such tasks in due time.\\

We organize the work in the following fashion:
In Section~\ref{sec: singularities}, alongside a review of geometric engineering with CY3 and CY4, we present general insights on singular Calabi-Yau spaces, drawing from the mathematical literature. 
After covering the more familiar case of terminal threefold singularities in Section~\ref{sec: terminal 3folds}, we move on to describe terminal fourfold singularities in Section~\ref{sec: terminal sing}.
These do not admit crepant resolutions, or admit crepant resolutions with at most 2- and 4-cycles in the exceptional set. In Sections \ref{sec: curve singularities} and \ref{sec: non iso curve singularities} we also introduce two classes of terminal fourfold singularities admitting small resolutions with at most exceptional 2-cycles, that furnish the foundation for our physics-oriented investigations. In Section \ref{sec: tachyon} a review of the tachyon condensation formalism is presented. This provides the main technical tool for our computations. In Section~\ref{sec: 3d SCFTs} and Section~\ref{sec: non iso 3d SCFTs} we flesh out the 3d $\mathcal{N}=2$ supersymmetric theories engineered by our chosen CY fourfolds in concrete examples.
Finally, in Section~\ref{sec: conclusion} we gather and discuss some concluding remarks. Appendix~\ref{app:resolutions} collects the rigorous technique for computing the resolution of the fourfold singularities examined throughout the work.

\section{Singular CY spaces and geometric engineering } \label{sec: singularities}

\subsection{Canonical and terminal singularities}

In this section we aim at gathering a collection of mathematical results about singular Calabi-Yau fourfolds. We reckon that this is essential in order to pursue advancements in the understanding of their physical role.

As a starting point, we review the definition of the mathematical arena we explore in this work, namely canonical and terminal singularities.\\

\indent A normal and quasi-projective CY $n$-fold $X$ has a \textit{canonical singularity} if and only if the following conditions are satisfied \cite{reid1985young}:
\begin{itemize}
    \item Its canonical divisor $K_X$ is such that $rK_X$ is a Cartier divisor; $r$ is the index of the singularity and it is greater than or equal to 1.
    \item For every resolution $f: Y\rightarrow X$, that produces exceptional divisors $E_i$, it holds:
    \begin{equation}\label{canonical}
        K_Y = K_X + \sum_i a_i E_i \quad \text{with} \quad  a_i \geq 0.
    \end{equation}
\end{itemize}
\textit{Crepant resolutions} preserve the Calabi-Yau condition, i.e.\ $a_i = 0, \forall i$.\\
\indent In the case of Calabi-Yau $n$-folds $X$ built as hypersurface singularities in an affine space $\mathbb{C}^{n+1}$ admitting a $\mathbb{C}^*$-action, the canonicity condition can be made very explicit. In full generality, $X$ can be expressed as:
\begin{equation}
    F(x_1,\ldots,x_{n+1}) = 0 \quad \subset \mathbb{C}^{n+1}.
\end{equation}
Suppose that there exists a $\mathbb{C}^*$ acting on the $x_i$ with weights $\omega_{x_i}$, $i=1,\ldots,n+1$
\begin{equation}
    (x_1,\ldots,x_{n+1}) \rightarrow (\lambda^{\omega_{x_1}}x_1,\ldots,\lambda^{\omega_{x_{n+1}}}x_{n+1}),
\end{equation}
such that $F$ is quasi-homogeneous with weight $\omega_F$ under it. \textit{Then} a theorem by Reid (see \cite{reid1980canonical} (4.1), (4.2), (4.3)) guarantees that $F$ is a canonical singularity if and only if:
\begin{equation}\label{canonicity}
    \sum_i \omega_{x_i} > \omega_F.
\end{equation}
From the physical standpoint, in the case $n=4$ it is paramount to remark that the condition \eqref{canonicity} is equivalent to the requirement that the 3d theory engineered by M-theory on $X$ flows to an IR fixed point \cite{Gukov:1999ya}. In this context, the $\mathbb{C}^*$-action is identified with the $U(1)$ R-symmetry of the resulting theory. All the singularities we will examine in Section~\ref{sec: curve singularities} and \ref{sec: non iso curve singularities} satisfy \eqref{canonicity}, and are hence good candidates to construct theories that flow to a 3d SCFT in the deep IR. \\

\indent Going back to the general picture, \textit{terminal singularities} are defined as canonical singularities with $a_i >0$ in \eqref{canonical} for every resolution. This is equivalent to stating that terminal singularities do \textit{not} admit any crepant resolution with compact divisors in the exceptional set. Furthermore, the following theorem holds (see \cite{reid1985young} and Corollary 8.3.2 of \cite{ishii2018introduction}):\\

\indent \hypertarget{thm1}{\textbf{Theorem 1:}} if $X$ is a Calabi-Yau $n$-fold with at worst terminal singularities, then the codimension of the singular locus is greater than or equal to 3.\\

In particular, the above theorem implies that terminal CY3 are isolated points, and that terminal CY4 can be either points or curves.

\subsection{Geometric engineering on CY3}
\indent We now quickly recap the singular Calabi-Yau threefold case: the general argument due to \cite{Xie:2017pfl}, and also applied by \cite{Closset:2020afy,Closset:2020scj,Closset:2021lwy}\footnote{For further works exploring 5d SCFTs constructed as explicit complete intersection singularities in M-theory, see e.g.\ \cite{DeMarco:2023irn,Tian:2021cif,Mu:2023uws,Bourget:2023wlb}}, reads that every non-compact CY threefold with \textit{canonical} singularities engineers a 5d $\mathcal{N}=1$ SCFT $\mathcal{T}_{5d}$. The massless degrees of freedom in the 5d effective theory can be detected examining BPS M2-brane states, as well as zero-modes of the M-theory 3-form $C_3$. In order to compute them explicitly, it is convenient\footnote{Although not necessary. For recent progress on the study of theories geometrically engineered from F-theory directly from the singular phase of the CY, see e.g.\ \cite{Collinucci:2014taa}.} to perform a crepant resolution $f: Y \rightarrow X$ of the singular CY3 $X$,\footnote{The choice of resolution is of course not unique. Different choices are related by (possibly multiple) flop transitions.} where the singular limit recovers the SCFT phase. In such a way, a precise mapping between the geometric features of the resolved CY3 $\tilde{X}$ and 5d SCFT data can be established:
\begin{itemize}
    \item Compact divisors (4-cycles) correspond to Coulomb branch parameters. One can see this by reducing $C_3$ on the Poincaré duals of the compact divisors. Thus:
    \begin{equation}
        \text{rank}(\mathcal{T}_{5d}) = \dim H_4(Y,\mathbb{Z}).
    \end{equation}
    \item Non-compact divisors correspond to flavour symmetry parameters. The rank of the flavor symmetry can be counted noticing that elements in $H_2(X,\mathbb{Z})$ are dual, via Poincaré duality, to compact and non-compact 4-cycles. Hence:
    \begin{equation}
        \text{rank}(\text{flavor}(\mathcal{T}_{5d})) = \dim H_2(Y,\mathbb{Z})- \dim H_4(Y,\mathbb{Z}).
    \end{equation}
\end{itemize}
In a subset of canonical singularities, the exceptional compact divisors admit a common ruling\footnote{I.e.\ there exists a ruling of the compact divisors such that the intersection $C = S_1\cap S_2$ between two of them is a fiber for both $S_1$ and $S_2$. Hence $C$ can be shrunk while keeping the volumes of $S_1$ and $S_2$ finite.}: 
in this case a low-energy 5d gauge theory phase exists, that flows to the UV 5d SCFT in the singular geometric limit. If more than one such ruling is present, multiple 5d low-energy theories flow to the same 5d SCFT, and are thus said to be UV dual.\\
\indent Most notably, the 5d Coulomb branch %{[RV:E' vero? dovremo dare una ref per questo statement.]}
can be studied employing the crepantly resolved phase of $X$. In particular, the prepotential of the effective theory can be computed via the triple intersection numbers of the compact divisors.\\
\indent A strikingly different picture holds for the Higgs branch: it corresponds to deformations of the singularity $X$, that are encoded by normalizable 3-forms. Consequently, M2-branes wrapped on the dual 3-cycles produce instantons that must be taken into account in order to characterize the Higgs branch \cite{Ooguri:1996me}. 
%To tackle this difficulty, a variety of techniques have been developed in a flurry of recent work, including magnetic quivers produced via $(p,q)$-web duals, as well as analyses relying on reduction of the M-theory background to Type IIA \cite{Collinucci:2021ofd,DeMarco:2021try,DeMarco:2022dgh}.

\subsection{Geometric engineering on CY4}\label{GE CY4}
The clear-cut dictionary between geometrical and physical data that has emboldened the progress in the 5d geometric engineering setting, gets murkier and much less understood in the context of M-theory on CY4 with canonical singularities, that corresponds to 3d $\mathcal{N}=2$ theories $\mathcal{T}_{3d}$.\\
\indent Let us start by recalling the (mostly trivial) undisputed facts on M-theory geometric engineering on CY4. One can, mimicking the CY3 case, tentatively work in the crepantly resolved phase $f:Y \rightarrow X$ of a Calabi-Yau fourfold $X$, that in general produces compact, as well as non-compact, 6-cycles, 4-cycles and 2-cycles.
\begin{itemize}
    \item In a chosen resolved phase, the rank of the 3d theory is given by the number of compact divisors (6-cycles) produced by the crepant resolution. This comes from the reduction of $C_3$ on the 2-forms that are Poincaré duals of the 6-cycles. Hence:
    \begin{equation}
        \text{rank}(\mathcal{T}_{3d}) = \dim H_6(Y,\mathbb{Z}).
    \end{equation}
    Similarly, the reduction of $C_3$ on the Poincaré duals of non-compact 6-cycles corresponds to the generators of the flavour symmetry of $\mathcal{T}_{3d}$. These can be counted as:
    \begin{equation}\label{flavor 3d}
        \text{rank}(\text{flavor}(\mathcal{T}_{3d})) = \dim H_2(Y,\mathbb{Z}) - \dim H_6(Y,\mathbb{Z}).
    \end{equation}
    Notice that expression \eqref{flavor 3d} only accounts for the factors of the flavor symmetry that are openly visible in the singular (or resolved) geometry. In general, other than non-abelian completions in the IR, there might be further rank enhancement due to effects related to the $G_4$ flux.\footnote{E.g.\ consider the setting involving a CY4 with a singular compact curve, supporting ADE singularities. Blowing up the singularities and threading the $G_4$ flux on the resulting $\mathbb{P}^1$-fibered 4-cycles produces a spectrum of matter multiplets, with a flavor symmetry of suitable rank acting on them. Such symmetry is not visible from the presence of non-compact 6-cycles. We stress that these setups will \textit{not} occur in our main classes of fourfold singularities of Sections \ref{sec: 3d SCFTs} and \ref{sec: non iso 3d SCFTs}.}
    \item M2-branes wrapped on compact 2-cycles give rise to BPS particles in $\mathcal{T}_{3d}$. Pinpointing their representation under the 3d Lorentz group is crucial in order to understand to which supersymmetric multiplet they belong. 
\end{itemize}
The most relevant difference with respect to the threefold case is that the 3d theory engineered by a fourfold generically receives quantum corrections from M5-branes wrapped on compact divisors. This makes identifying its features a highly non-trivial task. Furthermore, it is unclear what is the physical role played by the 4-cycles in the resulting 3d theory. M2-branes wrapped on cycles in the second homology of the 4-cycles can give rise to BPS states: these are tightly related to BPS invariants encoded in the Gopakumar-Vafa (GV) and Donaldson-Thomas (DT) theories, as pioneered by \cite{Klemm:2007in}, and that have recently attracted a substantial deal of attention in the mathematical community. For a selection of the latest works in this area, see e.g.\ \cite{Cao:2014bca,Nekrasov:2017cih,Nekrasov:2018xsb,Cao:2018wmd,Kononov:2019fni,Cao:2019fqq,Cao:2019tnw,Cao:2020vce,Cao:2020otr,Cao:2020hoy,Bousseau:2020fus,cao2022k,monavari2022canonical,cao2023donaldson,Piazzalunga:2023qik,Nekrasov:2023nai,Liu:2024bgp}. In a related fashion, in these settings one must take great care in analyzing the impact of the four-form flux $G_4$. Indeed, different choices of the background $G_4$ can produce radically different spectra in the underlying effective theory \cite{Donagi:2008ca,Braun:2011zm,Krause:2011xj,Hayashi:2008ba,Grimm:2011fx}. In addition, M5-branes can wrap generic 4-cycles producing tensionless strings in 3d in the vanishing volume limit.\\
\indent Nonetheless, despite these hints, a thorough physical 3d interpretation of the 4-cycles in Calabi-Yau fourfolds is, at present, lacking.\\

\indent Slightly more can be said about the role of 2-cycles that do not sit inside 6- or 4-cycles. M2-branes wrapping these cycles give rise to BPS states in the resulting 3d theory, that can be identified with \textit{chiral multiplets}. We will provide a physics argument for this statement in later sections, after we have introduced an interesting class of Calabi-Yau fourfolds that precisely displays such 2-cycles.

\section{Terminal singular Calabi-Yau threefolds}\label{sec: terminal 3folds}
\indent Before delving into the CY4 case, we list some facts about terminal singularities in the more familiar CY threefolds. 
As we have previously mentioned, they are isolated points, and in general they might or might not admit a crepant resolution. If they do, it is a small resolution, which produces a collection of 2-cycles as exceptional set. A chief example of these singularities, often used as template in the physics community, is the conifold. It can be represented as a quadratic equation in $\mathbb{C}^4$:
\begin{equation}
    x^2 +y^2+z^2+w^2 =0 \quad \subset \mathbb{C}^4.
\end{equation}
As excellently reviewed in \cite{Candelas:1989js}, the conifold admits two crepant small resolutions related by a flop transition, that replace the singular point with a single $\mathbb{P}^1$. Wrapping a M2-brane on such $\mathbb{P}^1$ produces a hypermultiplet in the effective theory along the 5 directions transverse to the conifold, that becomes massless in the singular limit, when the volume of the 2-cycle goes to zero. It has been proven by Reid (see \cite{reid1985young}, Theorem 3.2) that \textit{all} terminal CY3 singularities are of \textit{compound Du Val} (cDV) form, which means that they are a one-complex-parameter deformation of Du Val singularities (also known as Kleinian or ADE singularities). Indeed, notice that the conifold can be seen as a family of~$A_1$ singularities deformed by the complex parameter $w$. In general, cDV singularities are hypersurfaces in $\mathbb{C}^4$ that take the form:
\begin{equation}
    P_{\mathfrak{g}}(x,y,z) + wh(x,y,z,w) = 0,
\end{equation}
with $h$ an holomorphic function and $P_{\mathfrak{g}}(x,y,z) = 0$ is the usual presentation of the ADE singularities:
\begin{equation}
\begin{array}{ll}
  P_{A_k}= &  x^2+y^2+z^{k+1} \:,  \\
P_{D_k}=  &  x^2+zy^2+z^{k-1} \:, \\
P_{E_6}= & x^2+y^3+z^4 \:, \\
P_{E_7}= & x^2+y^3+yz^3 \:, \\
P_{E_8}= & x^2+y^3+z^5 \:.\\
\end{array}
\end{equation}
Isolated cDV singularities admitting a quasi-homogeneous action have been classified in \cite{Wang:2015mra}, and the related 5d SCFTs have been extensively studied in \cite{Collinucci:2021ofd,DeMarco:2021try,DeMarco:2022dgh}\footnote{Their role is crucial also in F-theory constructions, as highlighted in \cite{Arras:2016evy}.}. In general, they give rise to 5d theories with empty Coulomb branch and a collection of hypermultiplets, that can be either charged or uncharged under some abelian continuous or discrete flavour group. This can be proven noticing that they admit small resolutions with a collection of $\mathbb{P}^1$'s as exceptional set: the analysis of the spin of localized M2-branes carried out in \cite{Witten:1996qb} shows that these give rise to 5d hypermultiplets. Charged states under some $U(1)$ flavor group correspond to M2-branes wrapped on the exceptional $\mathbb{P}^1$'s, while uncharged states come from M2-branes localized at non-resolvable singularities \cite{Braun:2014nva,Arras:2016evy}.\\

Why are terminal singularities relevant to the geometric engineering program? Their importance can be nicely summarized recalling a theorem by Reid (see \cite{reid1985young}, Theorem 3.2), that we quote verbatim:\\

\indent \textbf{Theorem 2:} if $X$ is a threefold with canonical singularities then there exists a crepant partial resolution $f: Y\rightarrow X$ where $Y$ has only terminal singularities.\\

\indent In other words, terminal threefold singularities can appear as leftover singularities of CY3, even after performing a crepant partial resolution that has inflated a collection of compact divisors. Of course, terminal singularities do not modify the Coulomb branch of the 5d theory resulting from M-theory geometric engineering, but in general affect its Higgs branch. Thus, it is paramount to characterize them in detail. We will come back to this point shortly, when dealing with the less familiar fourfold case.

\section{Terminal singular Calabi-Yau fourfolds}\label{sec: terminal sing}
In the preceding sections we have introduced the generic features of the crepant resolution of a Calabi-Yau fourfold $X$. We have seen that the presence of compact divisors in the resolved phase yields challenging issues that hinder a rigorous understanding of the 3d effective theory, due to instantonic corrections coming from wrapped M5-branes.\\
\indent Therefore, from this point on we choose to exclusively devote our attention to \textit{terminal fourfold singularities}, namely Calabi-Yau fourfolds that admit crepant resolutions with \textit{at most} 4- or 2-cycles. These play a role that is akin to the terminal \textit{threefold} singularities, namely they appear as leftover singularities in CY4, after crepant partial resolutions that produce 6-cycles as exceptional sets.

In this section we collect, in a physically-oriented mindset, the state of the art of the mathematical knowledge about terminal CY4, paying extra care about the features of their crepant resolutions.

As a first, it is useful to pictorially represent the overarching relations that organize Calabi-Yau fourfolds with terminal singularities.
Because of \hyperlink{thm1}{Theorem 1}, terminal Calabi-Yau fourfolds can possess either isolated singular points or singular curves. Besides, for each of these cases, one can construct terminal CY4 that are complete intersections, or that are non-complete intersections. As is natural to expect, more leverage is available for complete intersection cases, as well as for toric ones. See Table \ref{fig:diagram} for a visual recap of the possible cases: we will come back to it in due time, adding relevant information about the features of each class of singularities.\\

 \begin{table}[t!]
    \centering
    \scalebox{0.8}{\begin{tikzpicture}
    \draw[rounded corners,thick] (0, 0) rectangle (18, 6) {};
    \node at (9,6.3) {CY4 terminal singularities};
    \node at (4.5,4.5) {Isolated \& Complete Intersection};
    \node at (13.5,4.5) {Isolated \& Non-complete Intersection};
    \node at (4.5,1.5) {Non-isolated \& Complete Intersection};
    \node at (13.5,1.5) {Non-isolated \& Non-complete Intersection};
    \draw[thick] (0,3)--(18,3);
    \draw[thick] (9,0)--(9,6);
    \end{tikzpicture}}\;
    \caption{Schematic summary of terminal CY4.}
    \label{fig:diagram}
    \end{table}

\subsection{Terminal isolated singular Calabi-Yau fourfolds}

We start by dealing with isolated terminal CY4. 

\subsubsection{Complete intersection case}

The analysis of terminal isolated singular CY4 is relatively straightforward in the complete intersection case. Indeed, the following theorem holds \cite{ishii2018introduction}:\\

\indent \hypertarget{thm3}{\textbf{Theorem 3:}} if $X$ is a complete intersection CY4 with terminal isolated singularities, then it takes one of the two following forms:
\begin{itemize}
    \item $X$ is a hypersurface in $\mathbb{C}^5$. Furthermore, it is either a double or triple point.
    \item $X$ is a complete intersection of two double hypersurfaces in $\mathbb{C}^6$.
\end{itemize}
An explicit classification of the above cases is, to the best of our knowledge, still missing. Nonetheless, their resolutions can be completely characterized, thanks to Samuel's conjecture, proven by Grothendieck in \cite{grothendieck2005cohomologie}:\\

\indent \hypertarget{thm4}{\textbf{Theorem 4:}} if $X$ is a complete intersection CY4 with terminal isolated singularities, then it \textit{does not} admit any small resolution.\\

At first sight, Theorem 4 seems to imply that complete intersection CY4 with terminal isolated singularities bear no interesting physical content, as there is no 2-cycle along which M2-branes can be wrapped. As the threefold case has amply demonstrated \cite{Collinucci:2021wty,Collinucci:2021ofd}, though, this is not entirely correct. Indeed, M2-branes can still give rise to uncharged chiral multiplets in 3d, as we will concretely see in Section~\ref{sec: 3d SCFTs}.

\subsubsection{Non-complete intersection case}

\begin{figure}[H]
    \centering
             \centering
    \scalebox{0.9}{\begin{tikzpicture}
    \draw[fill=black] (0,0) circle (0.05);
    \draw[fill=black] (1,0) circle (0.05);
    \draw[fill=black] (1.4,0.7) circle (0.05);
    \draw[fill=black] (0.4,0.7) circle (0.05);
    \draw[fill=black] (0.4,1.7) circle (0.05);
    \draw[fill=black] (1,-1) circle (0.05);
    \draw (0,0)--(1,0)--(1.4,0.7);
    \draw[dashed] (1.4,0.7)--(0.4,0.7)--(0,0);
    \draw (0.4,1.7)--(0,0);
    \draw (0.4,1.7)--(1,0);
    \draw (0.4,1.7)--(1,0);
    \draw (0.4,1.7)--(1.4,0.7);
    \draw[dashed] (0.4,1.7)--(0.4,0.7);
    \draw (1,-1)--(1,0);
    \draw (1,-1)--(1.4,0.7);
    \draw (1,-1)--(0,0);
    \draw[dashed] (1,-1)--(0.4,0.7);
    \node at (0.5,-1.6) {(a)};
    %%%%%
    \draw[fill=black] (5,0) circle (0.05);
    \draw[fill=black] (6,0) circle (0.05);
    \draw[fill=black] (6,0.7) circle (0.05);
    \draw[fill=black] (5,1.4) circle (0.05);
    \draw[fill=black] (6,-1) circle (0.05);
    \draw (5,1.4)--(5,0);
    \draw (5,1.4)--(6,0);
    \draw (5,1.4)--(6,0.7);
    \draw (5,0)--(6,0)--(6,0.7);
    \draw[dashed] (5,0)--(6,0.7);
    \draw (6,-1)--(5,0);
    \draw (6,-1)--(6,0);
    \draw[dashed] (6,-1)--(6,0.7);
    \node at (5.5,-1.6) {(b)};
    \end{tikzpicture}}\;
    \caption{Toric fans for examples of terminal isolated toric CY4. The resolved phase of case (a) is  $\mathcal{O}(-1,-1)\oplus\mathcal{O}(-1,-1)\rightarrow \mathbb{P}^1\times \mathbb{P}^1$, for case (b) is $\mathcal{O}(-1)\oplus\mathcal{O}(-2)\rightarrow \mathbb{P}^2$.}
    \label{fig:toric cases}
    \end{figure}
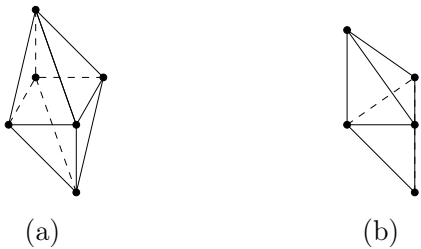

\indent Samuel's conjecture can be easily escaped by considering non-complete intersection terminal isolated CY4. These generically admit a small resolution, and the simplest examples can be drawn from the zoo of toric geometry. A toric CY4 is described by a fan in $\mathbb{Z}^4$, whose elements have endpoints on a common 3d sublattice. In order for the singularity to be isolated, the toric diagram cannot have parallel edges along the same line, as these would give rise to surfaces of singularities, as well as external non-triangular faces, because they would produce lines of singularities corresponding to the shape of the face, if the latter is interpreted as a toric diagram for a CY3.\footnote{E.g.\ a square face produces a line of singularities of conifold type.} Thus, one is left with possibilities such as the local cone over $\mathbb{P}^1\times \mathbb{P}^1$, or over $\mathbb{P}^2$, that we represent in Figure \ref{fig:toric cases}(a) and \ref{fig:toric cases}(b), respectively.

By construction, these cases admit a small crepant resolution that produces a 4-cycle as exceptional set (a $\mathbb{P}^1\times \mathbb{P}^1$ or a $\mathbb{P}^2$, respectively).\\
\indent Further examples of terminal fourfold singularities that \textit{are not} complete intersections can be displayed employing orbifold singularities of the type $\mathbb{C}^4/G$, with $G \subset SL(4,\mathbb{C})$. In particular, Theorem 2.1 of \cite{anno2003four} shows that choosing $G$ as a suitable cyclic group, the corresponding orbifold singularity is isolated and terminal. It is also non-complete intersection, thanks to a Theorem by \cite{dais1998all}, which proves:\\

\indent \hypertarget{thm5}{\textbf{Theorem 5:}} all CY4 of the form $\mathbb{C}^4/G$, with $G$ an abelian subgroup of $SL(4,\mathbb{C})$, and that can be written as complete intersections in some affine space, are not terminal.\\

There exist also 
non-abelian subgroups of $SL(4,\mathbb{C})$ that produce terminal isolated singular fourfolds \cite{anno2003four}, although it is still unclear whether they can be represented as a complete intersection.

\subsubsection*{Final remarks on isolated terminal CY4}
\indent We have seen that complete intersection cases do not admit a small resolution, and that non-complete intersection cases can admit a small resolution with 4-cycles as exceptional set.\footnote{We are not aware of any example of non-complete intersection fourfold singularity that does not admit any resolution, even a small one. %They might not exist, or simply might not have been spotted yet, given the focus of most on the literature on crepant resolutions.
} The following question then naturally arises: do terminal isolated CY4 admitting a small resolution with only 2-cycles in the exceptional set exist? It turns out that the answer is negative, thanks to a theorem dating back to the work of Laufer\footnote{We thank Andreas P. Braun for bringing this result to our attention.} \cite{laufer1981cp1}:\\

\indent \hypertarget{thm6}{\textbf{Theorem 6:}} if $X$ is a CY $n$-fold with an isolated singularity at the point $P$, admitting a small crepant resolution $f: Y\rightarrow X$, such that $f^{-1}(P)$ is one-dimensional and irreducible, then $n=3$ and $f^{-1}(P)$ is a $\mathbb{P}^1$.\footnote{A slightly relaxed version of the theorem that does not require irreducibility, predicts that the exceptional fiber can be a collection of $\mathbb{P}^1$'s, as it happens in a plethora of isolated cDV threefold cases.}\\

Laufer also predicted the normal bundles of said $\mathbb{P}^1$, thus classifying simple flops. For our aims, Theorem 6 excludes the existence of terminal isolated CY4 with only 2-cycles in the exceptional set. Let us summarize the main features of terminal isolated CY4 in Table \ref{tab:isolated summary}.

\renewcommand{\arraystretch}{1.4}
\begin{table}[ht]\centering
\begin{equation}
\scalemath{0.9}{
\begin{array}{|c|c|}
\hline
\hline
\multicolumn{2}{|c|}{\textbf{Isolated terminal CY4}} \\
\hline
\textbf{Complete intersection} & \textbf{Non-complete intersection}\\
\hline
 \text{Hypersurfaces in $\mathbb{C}^5$ or complete intersections in $\mathbb{C}^6$}
& \text{Orbifolds of $\mathbb{C}^4$, toric geometry, etc...}\\
\hline
\text{Never admit crepant resolutions} & \text{Some cases admit small crepant resolution with 4-cycles}\\
\hline
\hline
\end{array}}\nonumber
\end{equation}
\caption{Properties of isolated terminal CY4.}\label{tab:isolated summary}
\end{table}

\subsection{Terminal non-isolated singular Calabi-Yau fourfolds}
In analogy with the Calabi-Yau threefold case, much less is known about non-isolated fourfold singularities, and attempts at classifying them become even harder. In the fourfold case, \hyperlink{thm1}{Theorem 1} implies that non-isolated terminal singularities can appear only on curves, and not on surfaces. In order to understand the general features of these singularities, the following theorem by Reid provides useful resources:\\

\indent \hypertarget{thm7}{\textbf{Theorem 7:}} if $X$ is a CY $n$-fold with canonical singularities, then so has its general hyperplane section\footnote{Namely, the theorem holds for the intersection of $X$ with a generic hyperplane of the ambient space, except for possibly a finite set of hyperplanes. E.g. for the conifold case
$x^2 +y^2+z^2+w^2 = 0$,
the intersection with the ambient space hyperplane $w=0$ is a $A_1$ singularity. For further details, we refer to Definition 2.5 of \cite{reid1980canonical}.}.\\

Theorem 7 is equivalent to the statement that the CY $n$-fold locally looks like the Cartesian product of $\mathbb{C}$ with a canonical singularity of dimension $n-1$, \textit{except} possibly for a finite set of \textit{dissident points}  (see \cite{reid1980canonical}, Theorem 1.14), where the singularity can enhance.\\
\indent Let us flesh out the consequences of Theorem 7. For the $n=3$ case, this entails that up to dissident points the Calabi-Yau locally looks like $\mathbb{C} \times ADE$, as $ADE$ singularities are the only canonical singularities in dimension 2. The simplest non-trivial example one can think of is the intersection of two $A_1$ singularities:\footnote{Or, also, a non-compact line of $A_1$ singularities enhancing to $A_2$ on a dissident point, as in the threefold $x^2+y^2=z^2w$. We stick to the example in the text for the sake of symmetry between the two lines, although it is by no means necessary. We will follow the same philosophy later in the text, when discussing the CY4 case.}
\begin{equation}\label{spp}
  X: \quad  x^2+y^2 = z^2w^2 \quad\subset \mathbb{C}^4.
\end{equation}
We see that \eqref{spp} has two singular lines at:
\begin{equation}\label{singular lines}    x = y = z = 0, \quad\quad x = y = w = 0.
\end{equation}
Notice that, along the singular lines \eqref{singular lines}, but far from the intersection point $x= y = z = w = 0$, the threefold looks like $\mathbb{C}\times A_1$. Indeed, e.g. along the line $x= y = z = 0$ the local description of the threefold reads:
\begin{equation}
    X: \quad\{ x^2+y^2 = z^2 \}\times \mathbb{C}_w.
\end{equation}
Near the origin, instead, the singularity enhances to $A_3$, as both $z$ and $w$ are infinitesimals:
\begin{equation}
  X: \quad  x^2+y^2= z^2w^2|_{z\sim \lambda, w\sim \lambda} \quad \rightarrow \quad x^2+y^2 = \lambda^4.
\end{equation}
As is well-known, the 5d SCFT engineered from M-theory on \eqref{spp} is composed of hypermultiplets charged under the flavor $\mathfrak{su}(2)\oplus\mathfrak{su}(2)$ \cite{Katz:1996xe}.\\

\indent For the $n=4$ case, \hyperlink{thm7}{Theorem 7} has two consequences:
\begin{itemize}
    \item if the singular locus is a line, then locally the fourfold looks like $\mathbb{C} \times \text{canonical CY3}$, except for dissident points.
    \item if the singular locus is a surface, then locally the fourfold looks like $\mathbb{C}^2 \times \text{Du Val}$, except for dissident lines and points;\footnote{This can be easily seen e.g.\ considering a canonical CY4 with lines of singularities supporting canonical CY3 that are also singular along lines, thus producing surface singularities in the CY4. Applying \hyperlink{thm7}{Theorem 7} to such canonical CY3 then shows that the fiber of the surface singularity in the CY4 is necessarily a Du Val singularity, apart form dissident lines and points.}
\end{itemize}
In this work, we are especially interested in studying terminal CY4 singularities. We will hence concentrate on the case in which the CY4 admits at most singular lines, further requiring that they support terminal CY3. 
%According to \hyperlink{thm6}{Theorem 6}, such setup ensures that non-compact 4-cycles can arise, with additional isolated dissident points. 
We will show that in the classes of singularities that we will introduce in Sections \ref{sec: curve singularities} and \ref{sec: non iso curve singularities} this \textit{automatically guarantees} that the resulting CY4 is terminal.\\
\indent It is worth pointing out that this approach enables us to harness the knowledge accumulated in recent years about canonical CY3 in order to construct novel examples of terminal CY4. Of course, our aim in studying terminal CY4 is to construct physically interesting 3d $\mathcal{N}=2$ theories via M-theory geometric engineering. Therefore, when working with non-isolated singularities, we wish to focus on cases that display an \textit{intersection} of at least two lines of singularities, following precisely the same philosophy of the example \eqref{spp}. In such fashion, one builds a terminal CY4 with at least two intersecting singular lines, that locally looks like $(\mathbb{C}\times \text{canonical CY3})$ far from the intersection point, and displays a dissident point at the intersection, where a non-trivial 3d $\mathcal{N}=2$ theory might lie.\\

This is the appropriate moment to update our schematic summary of CY4 terminal singularities, collecting their features in Table \ref{fig:diagram2}.

 \begin{table}[H]
    \centering
    \scalebox{0.8}{\begin{tikzpicture}
    \draw[rounded corners,thick] (0, 0) rectangle (20, 6) {};
    \node at (10,6.3) {\textbf{CY4 terminal singularities}};
    \node at (5,4.9) {\textbf{Isolated} \& \textbf{Complete Intersection}};
    \node at (5,4.1) {No small crepant resolution };
    \node at (15,4.9) {\textbf{Isolated} \&\textbf{ Non-complete Intersection}};
    \node at (15,4.1) {Might admit small crepant resolution};
    \node at (15,3.66) {with exceptional 4-cycles};
    \node at (5,1.9) {\textbf{Non-isolated} \& \textbf{Complete Intersection}};
    \node at (5,1.1) {Might admit small crepant resolution};
    \node at (5,0.65) {with exceptional 2-cycles and 4-cycles};
    \node at (15,1.5) {\textbf{Non-isolated} \& \textbf{Non-complete Intersection}};
    \draw[thick] (0,3)--(20,3);
    \draw[thick] (10,0)--(10,6);
    \end{tikzpicture}}\;
    \caption{Updated schematic summary of terminal CY4.
    }
    \label{fig:diagram2}
    \end{table}

\indent In the next two sections, we finally reap the fruits of this general recap focused on terminal Calabi-Yau fourfolds, introducing two classes of terminal CY4, respectively isolated and non-isolated, that are amenable to a thorough physical analysis. We proceed in steps of increasing intricacy: first we introduce a class of isolated singularities that admit \textit{no} small resolution, and then a class of non-isolated singularities allowing a small resolution with a collection of $\mathbb{P}^1$'s as exceptional set.

\subsection{A class of isolated singular Calabi-Yau fourfolds}\label{sec: curve singularities}
In this section, we introduce a class of fourfold terminal singularities that are prone to a detailed physical investigation. A subset of these appeared in the pioneering work of \cite{Gukov:1999ya}, even though within a different context. Another single example was studied in the mathematical literature in \cite{kachi1996flips} with eyes on flip transitions. In this work, we introduce a full characterization of this class and examine it from a novel physical perspective.\\
\indent In order to define our class of terminal isolated CY4, we start from the well-known cDV threefold case, that we have briefly reviewed in Section~\ref{sec: terminal 3folds}. The isolated quasi-homogeneous cDV threefolds are hypersurfaces in $\mathbb{C}^4$ built as one-parameter deformations of a Du Val singularity labelled by the Lie algebra $\mathfrak{g} \in ADE$:
\begin{equation}
    \text{cDV threefolds:} \quad P_{\mathfrak{g}}(x,y,z) +wh(x,y,z,w)=0,
\end{equation}
that admit a quasi-homogeneous action with weights $(\omega_x,\omega_y,\omega_z,\omega_w)$:
\begin{equation}
    (x,y,z,w) \rightarrow (\lambda^{\omega_x}x,\lambda^{\omega_y}y,\lambda^{\omega_z}z,\lambda^{\omega_w}w).
\end{equation}
These have been completely classified in \cite{Wang:2015mra}: it turns out that only a finite number of families, each comprising infinite inequivalent singularities, exist. A notable example is the Reid pagoda \cite{reid1983minimal}, that reduces to the conifold in the $k=1$ case:
\begin{equation}
    x^2+y^2+z^{2k}+w^2=0.
\end{equation}

It is natural to ponder whether a similar classification is achievable by considering \textit{two-parameter} complex deformations of ADE singularities. Of course, this would produce a Calabi-Yau fourfold. Hence, we wish to classify \textit{quasi-homogeneous isolated singularities} of the form:
\begin{equation}\label{two par def}
\text{ccDV fourfolds:}\quad    P_{\mathfrak{g}}(x,y,z) +h(x,y,z,w,t)=0,
\end{equation}
where $w$ and $t$ are complex parameters, and $h$ is a holomorphic function satisfying $h(x,y,z,0,0)=0$. Abusing notation, and for the sake of briefness, one might baptize them {\it compound-compound-Du-Val (ccDV)} fourfolds. The $\mathbb{C}^*$ quasi-homogeneity acts as:
\begin{equation}\label{C star weights}
    (x,y,z,w,t) \rightarrow (\lambda^{\omega_x}x,\lambda^{\omega_y}y,\lambda^{\omega_z}z,\lambda^{\omega_w}w,\lambda^{\omega_t}t),
\end{equation}
and leaves the hypersurface \eqref{two par def} invariant up to a scale. For the sake of concreteness, the simplest example one can construct is a one-parameter deformation of the conifold:
\begin{equation}
x^2+y^2+z^2+w^2+t^2=0 \quad \subset \mathbb{C}^5,
\end{equation}
which is manifestly isolated.\\
\indent The most efficient method to classify quasi-homogeneous ccDV isolated singularities has been introduced by \cite{arnold2012singularities} and reviewed by \cite{Wang:2015mra}. 
With no loss of generality, we can set the coordinate system such that the isolated singularity lies at the origin $x=y=z=w=t=0$. The algorithm of \cite{arnold2012singularities} allows to determine the necessary conditions for a hypersurface equation of the form \eqref{two par def} to possess a isolated singularity at the origin. The criterion is as follows: in order \textit{not} to produce a singularity along the $x_i$ line, with $x_i$ one among $x,y,z,w,t$, the hypersurface \eqref{two par def} must possess at least one monomial of the form:
\begin{equation}\label{arnold condition}
    \prod_j x_j^{k_j}, \quad \text{with } \sum_j k_j -k_i\leq 1.
\end{equation}
It turns out that the condition \eqref{arnold condition} is extremely constraining. Applying it carefully, it is straightforward to classify all inequivalent\footnote{As it happens in the threefold case, a singularity from a family may be equivalent to one of another family. Crucially, though, \textit{not all} the singularities from a family can be interpreted as singularities from other families, thus making the two classes sensibly distinct.} quasi-homogeneous isolated singularities of the form \eqref{two par def}, that are collected in Table \ref{ccDV table}.  It is relevant to point out, as recalled by \cite{Gukov:1999ya}, that a Theorem by Tian and Yau \cite{tian1991complete} guarantees that the isolated ccDV singularities in Table \ref{ccDV table} possess an asymptotically conical Calabi-Yau metric.

\renewcommand{\arraystretch}{1}
\begin{table}[H]\centering
\begin{equation}
\scalemath{0.85}{
\begin{array}{|c|c|c|}
%\Xhline{4\mathbb Crrayrulewidth}
\hline 
 \textbf{ADE algebra} &  \textbf{CY4 singularity}  \\
\hline
\hline
\multirow{5}{*}{$A_{n}$} & x^2+y^2+z^{n+1}+w^k+t^j = 0 \\
 & x^2+y^2+z^{n+1}+z w^k+t^j = 0 \\
  & x^2+y^2+z^{n+1}+z w^k+w t^j = 0  \\
   & x^2+y^2+z^{n+1}+t w^k+t^j = 0  \\
 & x^2+y^2+z^{n+1}+tw^k+wt^j = 0 \\
\hline
\multirow{5}{*}{$D_{n}$} & x^2 +z y^2+z^{n-1}+w^k+t^j =0 \\
& x^2 +z y^2+z^{n-1}+yw^k+t^j =0  \\
& x^2 +z y^2+z^{n-1}+yw^k+wt^j =0  \\
& x^2 +z y^2+z^{n-1}+w^k+wt^j =0 \\
& x^2 +z y^2+z^{n-1}+tw^k+wt^j =0 \\

\hline
\multirow{8}{*}{$E_6$} & x^2 + y^3+z^4+w^k+t^j =0 \\
& x^2 + y^3+z^4+z w^k+t^j =0 \\
& x^2 + y^3+z^4+y w^k+t^j =0 \\
& x^2 + y^3+z^4+yw^k+zt^j =0  \\
& x^2 + y^3+z^4+tw^k+wt^j =0  \\
& x^2 + y^3+z^4+w^k+w t^j =0  \\
& x^2 + y^3+z^4+z w^k+w t^j =0  \\
& x^2 + y^3+z^4+y w^k+w t^j =0 \\
\hline
\multirow{5}{*}{$E_7$} & x^2+y^3+yz^3+w^k+t^j=0  \\
& x^2+y^3+yz^3+z w^k+t^j=0 \\
& x^2+y^3+yz^3+zw^k+wt^j=0\\
& x^2+y^3+yz^3+w^k+wt^j=0  \\
& x^2+y^3+yz^3+tw^k+wt^j=0  \\
\hline
\multirow{8}{*}{$E_8$} & x^2 + y^3+z^5+w^k+t^j =0  \\
& x^2 + y^3+z^5+z w^k+t^j =0 \\
& x^2 + y^3+z^5+y w^k+t^j =0  \\
& x^2 + y^3+z^5+yw^k+zt^j =0  \\
& x^2 + y^3+z^5+tw^k+wt^j =0 \\
& x^2 + y^3+z^5+w^k+w t^j =0 \\
& x^2 + y^3+z^5+z w^k+w t^j =0 \\
& x^2 + y^3+z^5+y w^k+w t^j =0 \\
\hline
\end{array}}\nonumber
\end{equation}
\caption{Classification of quasi-homogeneous isolated ccDV fourfold singularities. $n, k, j$ are non-negative integers.}\label{ccDV table}
\end{table}

Notice that, despite being much richer than the threefold case, the ccDV fourfolds still come in a finite number of families (each with infinite inequivalent singular family members). Furthermore, by virtue of \hyperlink{thm4}{Theorem 4}, they never admit a small resolution. We remark a relevant fact: despite carving out a large subset of fourfold singularities admitting no crepant resolution, the quasi-homogeneous ccDV isolated fourfolds do not exhaust all the possible cases. Indeed, as a consequence of \hyperlink{thm3}{Theorem 3}, one can build, e.g., triple points that cannot be crepantly resolved, such as the famous example exhibited by Reid in \cite{reid1985young}:
\begin{equation}
    x^3+y^4+z^4+w^6+t^6=0.
\end{equation}

\indent We will make clear the relevance of ccDV singularities for the physics of 3d $\mathcal{N}=2$ theories in Section~\ref{sec: 3d SCFTs}.
Before delving into the applications, in the next section we introduce a class of non-isolated fourfold terminal singularities, that is equally amenable to a physics-oriented analysis.

\subsection{A class of non-isolated singular Calabi-Yau fourfolds with exceptional curves}\label{sec: non iso curve singularities}

In the previous section we have introduced a class of fourfold singularities that can be thought of as the natural generalization to the fourfold setting of the conifold and of quasi-homogeneous cDV isolated threefold singularities. We have also seen that these Calabi-Yau fourfolds do not admit any crepant resolution. The next natural step is to ponder whether we can construct any singular CY4 that admits crepant resolutions with only 2-cycles in the exceptional set. Physically, these singularities would be particularly appealing: no instantonic corrections appear, because no compact 6-cycles are there, and no subtlety due to the $G_4$ flux arises, because of the lack of compact 4-cycles. At the same time, interesting physics can sprawl from wrapping M2-branes on the 2-cycles.\\
\indent As we have recalled in \hyperlink{thm6}{Theorem 6}, no singular isolated CY4 giving rise to an exceptional 2-cycle can exist. Hence, we are forced to recur to \textit{non-isolated} singular CY4. The inspiration to build the appropriate CY4 comes from the non-isolated threefold cases, such as the suspended pinch point:
\begin{equation}\label{spp2}
    x^2+y^2 =z^{k} w^{k},
\end{equation}
where $k$ is an integer. 
Notice that an alternative way to think of \eqref{spp2} is as a base-change applied to a $A_{k-1}$ singularity:
\begin{equation}\label{spp base change}
\begin{cases}
     x^2+y^2 = a^k\\
     a = zw\\
\end{cases}.
\end{equation}
It is easy to see that \eqref{spp base change} displays two non-compact lines of singularities of type $A_{k-1}$ along $x=y=z=0$ and $x=y=w=0$. At the intersection point the singularity enhances to $A_{2k-1}$. In a local patch far from the intersection point, the singularity looks like $\mathbb{C} \times A_{k-1}$, and therefore the exceptional set of its small crepant resolution is identical to the exceptional set of $A_{k-1}$, fibered over a complex line. The only non-trivial fiber of the resolution of \eqref{spp base change} can occur on top of the origin. A suitable small resolution is easily performed blowing up a Weil non-Cartier divisor, and it produces a collection of $2k-1$ $\mathbb{P}^1$'s arranged like the $A_{2k-1}$ Dynkin diagram on the fiber above the origin.\\

\begin{figure}[H]
    \centering
   \centering
 \scalebox{0.9}{
    \begin{tikzpicture}
        \draw[thick] (-2.5,-3)--(2.5,3);
        \draw[thick] (-2.5,3)--(2.5,-3);

        \draw[thick,dashed,->] (-1,3.5) to [out=270, in = 40] (-1.6,2.2);
        \draw[thick,dashed,->] (1,3.5) to [out=270, in = 140] (1.6,2.2);
        \draw[thick,dashed,<-] (0.5,0)--(1.5,0);
        \draw[fill=red] (0,0) circle (0.13);
        \node at (-1,4.8) {$\overbrace{\hspace{1.3cm}}^{\text{smooth}}$};
        \draw (-1,4.1) circle (0.3);
        %\draw[thick] (-3.2,3.8)--(-3.0,3.8);
        %\draw (-2.8,3.8) circle (0.15);
        %\draw[thick] (-2.6,3.8)--(-2.4,3.8);
        %\draw (-2.2,3.8) circle (0.15);
        %\draw[thick] (-2,3.8)--(-1.8,3.8);
        %\draw (-1.6,3.8) circle (0.15);
        %\draw[thick] (-1.4,3.8)--(-1.2,3.8);
        %\draw (-1,3.8) circle (0.15);
        %\draw[thick] (-2.2,4.0)--(-2.2,4.2);
        %\draw (-2.2,4.4) circle (0.15);
        %
        \node at (1,4.8) {$\overbrace{\hspace{1.3cm}}^{\text{smooth}}$};
        \draw (1,4.1) circle (0.3);
        %\draw[thick] (3.2,3.8)--(3.0,3.8);
        %\draw (2.8,3.8) circle (0.15);
        %\draw[thick] (2.6,3.8)--(2.4,3.8);
        %\draw (2.2,3.8) circle (0.15);
        %\draw[thick] (2,3.8)--(1.8,3.8);
        %\draw (1.6,3.8) circle (0.15);
        %\draw[thick] (1.4,3.8)--(1.2,3.8);
        %\draw (1,3.8) circle (0.15);
        %\draw[thick] (2.2,4.0)--(2.2,4.2);
        %\draw (2.2,4.4) circle (0.15);
        %
          \node at (2.2,-0.9) {$\underbrace{\hspace{1.3cm}}_{\text{singular}}$};
        \draw (2.2,0) circle (0.3);
        \draw[fill=black] (2.2,0.3) circle (0.06);
        \draw[fill=black] (2.2,-0.3) circle (0.06);
        %\draw[thick] (4.2,0)--(4,0);
        %\draw (3.8,0) circle (0.15);
        %\draw[thick] (3.6,0)--(3.4,0);
        %\draw (3.2,0) circle (0.15);
        %\draw[thick] (3,0)--(2.8,0);
        %\draw (2.6,0) circle (0.15);
        %\draw[thick] (2.4,0)--(2.2,0);
        %\draw (2,0) circle (0.15);
        %\draw[thick] (3.2,0.2)--(3.2,0.4);
        %\draw (3.2,0.6) circle (0.15);
        %
        \node[rotate=49] at (-1.5,-2.2) {$\boldsymbol{w=0}$};
        \node[rotate=-49] at (1.5,-2.2) {$\boldsymbol{t=0}$};
        \end{tikzpicture}}
    \caption{Small partial crepant resolution of \eqref{weyl eq}. Notice that outside the intersection point the resolved fourfold is smooth. At the intersection, there remain two singular non-resolvable points located on the north and south pole of the partially resolved fiber.}
    \label{fig:small res}
    \end{figure}
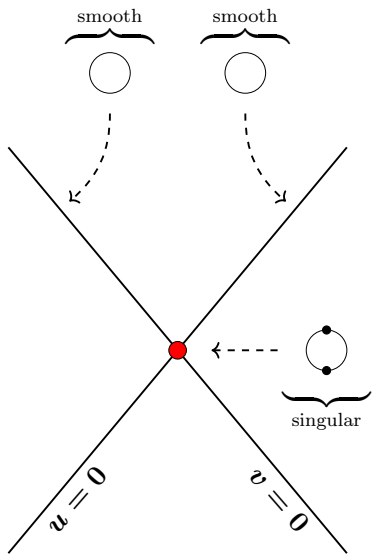

\indent In a completely analogous fashion, we can construct a CY4 with non-isolated singularities that intersect at a point, where the singularity is possibly enhanced. This time, as we wish to produce only 2-cycles in the exceptional set, the starting point is not a Du Val singularity, but rather a cDV singularity. Let us clarify this statement with an example. We can start from a cDV singularity, such as the famous Reid's pagoda \cite{reid1983minimal}, and apply a base-change à la \eqref{spp base change}:
\begin{equation}\label{spp 4fold}
\begin{cases}
     x^2+y^2 = z^{2k}-a^2\\
     a = wt\\
\end{cases}.
\end{equation}
Clearly, \eqref{spp 4fold} sports two lines of singularities at:
\begin{equation}
    x = y = z = w = 0, \quad x = y = z = t = 0.
\end{equation}
Locally, near a generic point of the singular lines, \eqref{spp 4fold} looks like $\mathbb{C}$ times the cDV threefold singularity $x^2+y^2 = z^{2k}-a^2$. Thus, a crepant resolution of these non-singular lines produces an exceptional set that is identical to the exceptional set of $x^2+y^2 = z^{2k}-a^2$, fibered over a complex line. At the intersection of the two singular lines, instead, the singularity may enhance.
Therefore, it is left to check that the dissident point of \eqref{spp 4fold} produces only 2-cycles as an exceptional set. We can perform the resolution explicitly. As a first step rewrite \eqref{spp 4fold} as
\begin{equation}\label{weyl eq}
    uv=(z^k+wt)(z^k-wt),
\end{equation}
and blow up the Weil non-Cartier divisor $u = z^k+wt=0$. In one of the charts the blow up looks like:
\begin{equation}
\begin{cases}
        u = \mu_1 \\
        z^k+wt = \mu_1\mu_2\\
        uv=(z^k+wt)(z^k-wt)\\
\end{cases}.
\end{equation}
Throwing out spurious equations the proper transform of \eqref{weyl eq} is then:
\begin{equation}\label{blowup eq}
    z^k+wt =\mu_1 \mu_2\\.
\end{equation}
Notice that, by a simple change of variables, \eqref{blowup eq} is precisely of the form:
\begin{equation}\label{isolated sing}
    x_1^2+x_2^2+x_3^k+x_4^2+x_5^2 = 0,
\end{equation}
which is an isolated quasi-homogeneous ccDV singularity, that we have extensively analyzed in the previous section, appearing in Table \ref{ccDV table}. As we have seen, the fourfold in \eqref{isolated sing} has an isolated singularity at the origin that admits no crepant resolution. In the other chart the picture is identical, namely we obtain another leftover singularity of type \eqref{isolated sing}.\\
\indent Let us summarize the result: the singular fourfold \eqref{weyl eq} admits a small partial resolution that completely resolves it outside the origin, where the exceptional set is a single $\mathbb{P}^1$. Thus, outside the origin the resolved fourfold \eqref{weyl eq} looks like $\mathbb{C}$ times the resolved threefold $x^2+y^2 = z^{2k}-a^2$. On top of the origin the small partial resolution inflates a $\mathbb{P}^1$ which is, however, not smooth. Indeed, two singular points of type \eqref{isolated sing} sit on top of the north and south poles. 
The pictorial result of the resolution is schematically represented in Figure \ref{fig:small res}.

All in all, the singularity \eqref{spp 4fold} cannot be completely resolved, as two points that admit no crepant resolution are left as a result of the small partial resolution. This will be key to keep in mind for correctly assessing the physical content of the 3d $\mathcal{N}=2$ theory arising from M-theory geometric engineering on \eqref{spp 4fold}.\\
\indent We plan on extracting it in the two upcoming sections.\\

\indent Before doing so, let us mention the obvious generalizations of \eqref{spp 4fold}: one can pick any $A$ singularity as a starting point, and apply a non-trivial base change. A more general case of type $A$ is:
\begin{equation}\label{general case}
    x^2+y^2 =z^k+w^{n_1}t^{n_2}.
\end{equation}
It has two lines of singularity on $x=y=z=w=0$ and $x=y=z=t=0$, and its resolvability depends on the specific choice of $n_1$ and $n_2$. In general, one expects that outside the origin the resolution looks like $\mathbb{C}$ times the resolved threefold that lies along the corresponding singular line. On the origin, additional singular isolated points of type $A$, classified in Table~\ref{ccDV table}, can appear. This shows a beautiful nested singularity structure that was absent in the threefold case: one can construct CY4 with non-isolated singularities that locally look like the product of a complex line with a canonical cDV threefold; at the intersection point of the non-compact lines, a phenomenon intrinsic to fourfold can arise, as nested isolated singularities of the type appearing in Table~\ref{ccDV table} can be leftover.\\

\indent A last option appears by moving away from the $A$ cases, namely considering constructions of type \eqref{spp 4fold} where the first equation is a cDV threefold built as a deformed $D$ or $E_6,E_7,E_8$ singularity. In this case, one might expect something fancier to happen at the dissident point, such as the appearance of higher-dimensional cycles in the exceptional set. It is easy to prove, though, that this is \textit{not} the case. Let us briefly sketch our argument here: all the singular CY4 we are constructing are ccDV singularities, namely they are devised as two-parameter deformations of some Du Val singularity labelled by $\mathfrak{g}\in ADE$. Therefore, they can be interpreted via the theory of Springer resolutions. For a review of its salient points, we refer to \cite{henderson2014singularities,yun2016lectures}. The key fact that we need here is that, thanks to a theorem by Brieskorn, every ccDV singularity can be seen as the intersection of the Slodowy slice through the subregular nilpotent orbit of $\mathfrak{g}$ with the fibers of the map that encodes the Casimirs of $\mathfrak{g}$, with a suitable change of basis acting on the Casimirs. Then, the Grothendieck-Springer theorem guarantees that the resolution of such singularity admits \textit{at most} exceptional $\mathbb{P}^1$'s, up to possible isolated non-resolvable singularities. In other words, the complex dimension of the Springer fibers is \textit{at most} one.\\
\indent The upshot of the preceding argument is that ccD and ccE Calabi-Yau fourfolds with two lines of non-isolated singularities intersecting at the origin behave analogously as ccA singularities of the same kind: they might admit a small resolution that inflates a collection of $\mathbb{P}^1$'s on top of every singular point, up to isolated singularities of the type appearing in Table \ref{ccDV table} sitting on the fiber above the origin. We outline a recipe for their crepant resolution in Appendix \ref{app:resolutions}, and leave an extended analysis of their features for future work.

\section{Tachyon condensation formalism}\label{sec: tachyon}
In order to concretely analyze the 3d $\mathcal{N}=2$ theories that arise from M-theory geometric engineering on ccDV fourfolds, we will make ample use of the formalism furnished by tachyon condensation \cite{Sen:1998sm,Donagi:2003hh,Collinucci:2014qfa,Collinucci:2014taa}. In this section, we quickly review its landmark features, with an eye to our needs for the subsequent sections.\\

We wish to map our M-theory configurations on a Calabi-Yau fourfold background to a Type IIA picture involving D6-branes. In order to do that, consider a stack of D8-branes equipped with a vector bundle $E$ and a stack of anti-D8-branes equipped with a vector bundle $F$. The bifundamendal strings stretching between the two stacks are encoded by the tachyon map $T$:
\begin{equation}\label{two term complex}
 \begin{tikzcd}
 F \arrow{r}{T} & E .
\end{tikzcd}
\end{equation}
Whenever $T$ acquires a non-trivial vev, tachyon condensation is kicked off: the D8 and anti-D8 stacks collide and annihilate, possibly leaving a D6-brane remnant that survives the annihilation. Such remnant lies along the cokernel of the map $T$:
\begin{equation}
 \begin{tikzcd}
0 \rar &  F \arrow{r}{T} & E \rar & \text{coker}(T) \rar & 0 .
\end{tikzcd}
\end{equation}
The simplest case is triggered if $T$ is the identity map: in this case the annihilation is complete and there is no remnant. If the profile of $T$ is less trivial, its cokernel is non-empty and denotes a sheaf that is supported over the D6-branes locus $\Delta$, defined as:
\begin{equation}
    \Delta: \quad \text{det}(T) = 0.
\end{equation}
In this case, it is convenient to denote the D6-brane locus via the two-term complex~\eqref{two term complex}.\\
\indent Naturally, the gauge bundles on the D8 and anti-D8-brane stacks induce a redundancy on the tachyon vev, acting from opposite sides:
\begin{equation} \label{BifundTransfT}
\begin{tikzpicture}[scale=2]
\node (E) at (0,0) {$F$};
\node[right = of E] (F)   {$E$};
\node[right = of F] (ar) {$\Longrightarrow$};
\node[right = of ar] (T) {$T$};
\node[right = of T] (Tprime) {$G_{\rm D8}\, \cdot T \cdot \, G_{\overline{\rm D8}}^{-1}\:.$};

 \draw (E) edge[->] node[above, font=\scriptsize]{$T$} (F);
 \draw (T) edge[->] (Tprime);
 \draw(E) edge[loop below] node[font=\scriptsize]{$G_{\overline{\rm D8}}$} (E) ;
 \draw(F) edge[loop below] node[font=\scriptsize]{$G_{\rm D8}$} (F) ;
\end{tikzpicture} 
\end{equation}
The gauge symmetry preserved on the D6-brane stack is given by the transformations in \eqref{BifundTransfT} that stabilize $T$.\\
\indent As first introduced in \cite{Sen:1998sm} and reviewed in \cite{Collinucci:2021wty} in the arena of cDV threefolds, we wish to study the spectrum of open strings stretching between the D6-branes. This will allow us to characterize the 3d $\mathcal{N}=2$ theory lying at the intersection of the branes. We refer to the above-mentioned works for the full details of this construction.\\
\indent From the point of view of the D6-brane complex, open strings modes are captured by $\text{Ext}^1(\text{coker}\hspace{0.05cm}T,\text{coker}\hspace{0.05cm}T)$. Diagramatically, this corresponds to the map $\delta T$ between the D6-brane complexes:
\begin{equation}
\begin{tikzcd}
& F \dar{\delta T } \rar{T} & E\\
F \rar{T} & E
\end{tikzcd}
\end{equation}
Taking into account the gauge redundancy, we obtain:
\begin{equation}\label{gauge diagram}
\begin{tikzcd}[row sep=huge, column sep=large]
& F \dlar[dashed]{G_{\overline{\rm D8}}} \dar{\delta T} \rar{T} & E \dlar[dashed]{G_{\rm D8}}\\
 F \rar{T} & E
\end{tikzcd}
\end{equation}
Therefore, the map $\delta T$ is defined up to linearized gauge transformations:
\begin{equation}\label{linear transformations}
\delta T \sim \delta T +T\cdot g_{\overline{\rm D8}} + g_{\rm D8} \cdot T\,.
\end{equation}

In the next section we put this technology into use, applying it to the case of ccDV Calabi-Yau fourfolds.

%\textcolor{purple}{RV: alla fine stiamo usando l'Higgs $\Phi$ in diverse parti, quindi forse vale la pena muovere le parti generali dalla sezione 6.4 a qua.}

\section{3d $\mathcal{N}=2$ theories from isolated Calabi-Yau fourfolds}\label{sec: 3d SCFTs}

In this section we set on exploring M-theory geometric engineering on the class of singularities introduced in Section~\ref{sec: curve singularities}: our main technical tool is the tachyon condensation formalism sketchily reviewed in Section~\ref{sec: tachyon}.

\subsection{Simple example of isolated CY4 singularity}

\indent Let us start with the an illustrative example, namely the fourfold hypersurface singularity
\begin{equation}\label{quadrifold}
    x^2+y^2+z^2+w^2+t^2=0 \quad \subset \quad \mathbb{C}^5.
\end{equation}
Notice that \eqref{quadrifold} is the simplest singular one-parameter deformation of the conifold. As we have seen, owing to \hyperlink{thm4}{Theorem 4}, it does not admit any crepant resolution. Nonetheless, we will momentarily show that the effective 3d $\mathcal{N}=2$ theory generated by M-theory on \eqref{quadrifold} displays a non-trivial physical content. In order to manifestly see so, let us describe \eqref{quadrifold} via tachyon condensation. As a first step, rewrite \eqref{quadrifold} applying a trivial change of variables:
\begin{equation}\label{C star fibration}
    u v = z^2+w^2+t^2.
\end{equation}
The form \eqref{C star fibration} is particularly helpful to highlight the following $\mathbb{C}^*$-action:
\begin{equation}\label{C star}
    (u,v,z,w,t) \rightarrow (\lambda u, \frac{v}{\lambda},z,w,t).
\end{equation}
The action \eqref{C star} allows us to take the Type IIA limit of M-theory on \eqref{C star fibration}, by reducing on the $S^1$ contained in the $\mathbb{C}^*$. The locus where this action degenerates dictates where the D6-branes in the Type IIA configuration lie, namely:
\begin{equation}\label{D6 brane}
    \Delta = z^2+w^2+t^2 = 0.
\end{equation}
It is crucial to remark that \eqref{D6 brane} is a single irreducible D6-brane, namely it admits no factorization in terms of simpler D6-branes. Nonetheless, the world-volume of the D6-brane is singular at a point in codimension 3 (complex) in $\mathbb{C}^3_{z,w,t}$. It is precisely at this special point that a 3d $\mathcal{N}=2$ theory dwells.\\
\indent We wish to discover whether any physical massless mode is localized at the singular point. To this end, it is particularly apt to describe the D6-brane \eqref{D6 brane} via the tachyon condensation technology.
Hence, we introduce a stack of 2 D8-branes and 2 $\overline{\text{D8}}$-branes, connected by the tachyon map specified by the complex:
\begin{equation}
 \begin{tikzcd}
 \mathcal{O}^{\oplus 2} \arrow{r}{T} & \mathcal{O}^{\oplus 2} .
\end{tikzcd}
\end{equation}
The tachyon map is explicitly defined as an element in $GL(2,\mathbb{C})$:
\begin{equation}\label{tachyon quadrifold}
    T = \left(\begin{array}{cc}
      z+iw   & t \\
       -t  & z-iw
    \end{array} \right).
\end{equation}
Notice that $T$ has a non-trivial cokernel, and thus the condensation of the D8-$\overline{\text{D8}}$ system leaves a D6-brane remnant on the locus:
\begin{equation}\label{D6 quadrifold}
    \Delta: \quad \text{det}(T) = z^2+w^2+t^2=0,
\end{equation}
which is precisely the brane locus encoded by the fourfold singularity \eqref{quadrifold}.
In order to extract the physical spectrum, we must consider fluctuations of the tachyon \eqref{tachyon quadrifold}, appropriately modding out by gauge redundancies, as summarized by the diagram \eqref{gauge diagram}. In our example, the fluctuations read:
\begin{equation}
\begin{tikzcd}[row sep=huge, column sep=large]
& \mathcal{O}^{\oplus 2} \dlar[dashed]{G_{\overline{\rm D8}}} \dar{\delta T} \rar{T} & \mathcal{O}^{\oplus 2} \dlar[dashed]{G_{\rm D8}}\\
 \mathcal{O}^{\oplus 2} \rar{T} & \mathcal{O}^{\oplus 2},
\end{tikzcd}
\end{equation}
with $G_{\rm D8}$ and $G_{\overline{\rm D8}}$ elements in $SL(2,\mathbb{C})$.
Effectively, this amounts to modding out by linearized gauge transformations $g_{D8}$ and $g_{\overline{D8}}$ as in \eqref{linear transformations}:
\begin{equation}\label{lin quadrifold}
\delta T \sim \delta T +T\cdot g_{\overline{\rm D8}} + g_{\rm D8} \cdot T\,,
\end{equation}
where in this specific example $g_{D8}$ and $g_{\overline{D8}}$ are generic elements of $\mathfrak{sl}(2,\mathbb{C})$:
\begin{equation}
    g_{D8} = \left(\begin{array}{cc}
      g_{11}   & g_{12} \\
       g_{21}  & -g_{11}
    \end{array} \right), \quad  g_{\overline{D8}} = \left(\begin{array}{cc}
      \overline{g}_{11}   & \overline{g}_{12} \\
       \overline{g}_{21}  & -\overline{g}_{11}
    \end{array} \right).
\end{equation}
The fluctuations of the tachyon can be parameterized (without loss of generality) as:
\begin{equation}
    \delta T = \left(\begin{array}{cc}
      \delta_1  & \delta_2+\delta_3 \\
       \delta_2-\delta_3  & -\delta_1
    \end{array} \right).
\end{equation}
Most importantly, notice that the elements $G_{D8}$ and $G_{\overline{D8}}$ that stabilize the tachyon are trivial:
\begin{equation}
    \text{Stab}(T) = \{\mathbb{1}\}.
\end{equation}
Namely, all the symmetry that acts on the D8-$\overline{\text{D8}}$ stack, and that would be interpreted as a flavor symmetry from the perspective of the theory sitting at the origin of the D6-brane locus, is broken once the tachyon profile \eqref{tachyon quadrifold} is triggered. This in turns implies that the 3d $\mathcal{N}=2$ theory arising from M-theory on \eqref{quadrifold} has no flavor symmetry.\\
\indent Let us identify the physical content of the 3d theory: explicitly modding out by the equivalence \eqref{lin quadrifold} it is easy to find that all fluctuations of the tachyon map are not localized at the singular point $u=v=z=w=t=0$, except:
\begin{equation}
   \delta_3 \sim \delta_3 +z(g_{12}+g_{21})+w(ig_{21}-ig_{12})+t(2g_{11}).
\end{equation}
It is clear that employing the redundancy offered by the complex parameters $g_{11},g_{12},g_{21}$, every dependence of the fluctuation $\delta_3$ on $z,w,t$ can be removed, effectively localizing it 
at the singular point $u=v=z=w=t=0$. Thus, a massless string mode is localized at the origin of the singular fourfold \eqref{quadrifold}. Lifting it to M-theory, we expect it to arise from a BPS M2-brane state localized at 
 the origin. It corresponds to a one-dimensional complex degree of freedom: this is the telltale sign that its role in the 3d $\mathcal{N}=2$ theory engineered by M-theory on \eqref{quadrifold} is played by a single \textit{chiral multiplet}. Furthermore, no scale is present in the geometry, and so it is natural to guess that the 3d $\mathcal{N}=2$ theory at hand  actually flows to a SCFT in the IR.\\
 \indent We can pictorially summarize the physical relevance of this introductory example as:
 \begin{figure}[H]
    \centering
             \centering
    \scalebox{0.9}{\begin{tikzpicture}
    \node at (0,0) {M-theory on};
    \node at (0,-0.6) {$x^2+y^2+z^2+w^2+t^2=0$};
    \node at (4,-0.3) {$\Longleftrightarrow$};
    \node at (8,0) {3d $\mathcal{N}=2$ theory of};
    \node at  (8,-0.6) {one uncharged chiral multiplet};
    %%%%%
    \end{tikzpicture}}\;
    \caption*{}
    \label{fig:summary quadrifold}
    \end{figure}
    \vspace{-1.4cm}
As one can readily notice by the fact that the singular point cannot be resolved in a crepant fashion, no flavor group is visible from the geometric perspective\footnote{As we have briefly reviewed in Section~\ref{GE CY4}, a non-trivial geometric flavor group requires a non-empty $H_2(\text{CY4},\mathbb{Z})$, in some resolved phase.}. This would lead us to conclude that the 3d $\mathcal{N}=2$ chiral multiplet is not free. For the time being, though, we cannot exclude subtler mechanisms of flavor enhancement that would indeed conspire to justify the presence of a full-fledged \textit{free} chiral multiplet in the spectrum. This phenomenon is familiar and completely akin to the case of M-theory geometric engineering on terminal CY3 \cite{Closset:2020scj,Collinucci:2021ofd}: these produce 5d $\mathcal{N}=1$ theories of hypermultiplets, enjoying a geometric flavor symmetry that is \textit{smaller} than what would be anticipated for free hypers. A satisfying explanation for both the CY3 and CY4 case is at present lacking. We will see this phenomenon recur in all the subsequent cases analyzed in this work.

\subsection{$A$ series}\label{sec: A series}
In this section we present the 3d $\mathcal{N}=2$ theories engineered by the fourfold singularities of type $A_n$ from Table~\ref{ccDV table}, generalizing the simple example \eqref{quadrifold}. For conciseness, we focus on the class:
\begin{equation}\label{A class}
    x^2+y^2+z^{n+1}+w^k+t^j = 0 \quad \subset  \quad \mathbb{C}^5.
\end{equation}
The fourfolds \eqref{A class} have a unique isolated singular point at the origin $x = y = z = w = t = 0$, and admit no crepant resolution.\\
\indent Without loss of generality, suppose that $n+1$ is greater or equal to $k$ and $j$. Then, we can interpret \eqref{A class} as a two-parameter deformation of a $A_n$ singularity, that can be rewritten in such a way to make the action of $\mathbb{C}^*$ manifest:
\begin{equation}\label{A class Cstar}
    uv = z^{n+1}+w^k+t^j.
\end{equation}
We wish to replicate the same recipe employed in the case of the previous section, namely:
\begin{itemize}
    \item Take the Type IIA limit of \eqref{A class Cstar} and detect the D6-brane locus.
    \item Via the tachyon condensation formalism, write down the explicit form of the tachyon for \eqref{A class Cstar}, embedding it in $GL(n+1,\mathbb{C})$.
    \item Compute fluctuations of the tachyon: the ones (if any) localized at the origin, correspond to genuine modes of the 3d $\mathcal{N}=2$ theory engineered by M-theory on \eqref{A class Cstar}. We will momentarily see that these modes correspond to uncharged chiral multiplets.
\end{itemize}
Let us show how this recipe works for a subclass of the singularities \eqref{A class}:
\begin{equation}
    x^2+y^2+z^{n+1}+w^2+t^2 = 0 \quad \subset \quad \mathbb{C}^5,
\end{equation}
labelled by $n$.\\
\indent According to \eqref{A class Cstar}, the D6-brane locus reads:
\begin{equation}
    \Delta_{n} = z^{n+1}+w^2+t^2=0.
\end{equation}
In order to compute genuine 3d zero-modes, we must find a tachyon map $T_{n}$ such that:
\begin{equation}
    \text{det}(T_{n}) = \Delta_{n}.
\end{equation}
This tachyon map can be explicitly written as the following $(n+1)\times(n+1)$ matrix:
\begin{equation}\label{iso tachyon}
    T_{n} = \left(\begin{array}{ccccccccc}
      z & 1 & 0 & &\cdots & & & & 0 \\
     0 & z & 1 & 0 & \cdots & & & & 0\\
       &  &  & \ddots & & & & & \\
    \vdots & \vdots & & z & w+it & 0 & & & \vdots \\
    & &  & & 0 & 1 & 0 & &\\
        &  &  &  &  &  & \ddots & &\\
      0 & & & &\cdots  & & z & & 1\\
       w-it & 0 &   & & \cdots& & & & z\\
    \end{array}\right).
\end{equation}
Notice that the tachyon map cannot be holomorphically diagonalized, nor can it even be reduced to a block diagonal form: this is the telltale sign that no crepant (even partial) resolution is admissible. This is of course in agreement with the much more general \hyperlink{thm4}{Theorem 4}. Physically, it is equivalent to observing that:
\begin{equation}
    \text{Stab}(T_n) = \{\mathbb{1}\}.
\end{equation}
Consequently, as in the example \eqref{quadrifold}, no flavor group is present in the 3d $\mathcal{N}=2$ theory engineered by $T_n$.\\
\indent It is crucial to remark that $T_{n}|_{z=w=t=0}$ belongs to the\footnote{The ``floor'' function takes as input a real number and yields the greatest integer that is less than or equal to it. ``Ceiling'' yields the smallest integer that is greater than or equal to it.} $[\text{ceiling}(\frac{n+1}{2}),\text{floor}(\frac{n+1}{2})]$  nilpotent orbit of $A_{n}$. This is needed in order to ensure that all the 3d zero-modes are accounted for.\footnote{Choosing a different tachyon, with the \textit{same} determinant and with $T_{n}|_{z=w=t=0}$ belonging to a \textit{different} nilpotent orbit is of course allowed by the tachyon condensation formalism. These correspond to T-brane states that localize a smaller number of 3d zero-modes. For further details on the structure of T-branes in the context of terminal threefold singularities see e.g. \cite{Collinucci:2021ofd,DeMarco:2021try}.}\\
\indent Once in possession of the explicit tachyon map, it is a matter of simple algorithmic effort to compute its fluctuations:
\begin{equation}
    \delta T \sim \delta T +T\cdot g_{\overline{\rm D8}} + g_{\rm D8} \cdot T\,,
\end{equation}
with $g_{\rm D8}$ and $g_{\overline{\rm D8}}$ generic elements in $\mathfrak{sl}(n+1,\mathbb{C})$. Running the machine one can find that the modes localized at the origin $x=y = z = w =t = 0$ are:
\begin{equation}\label{3d modes A class}
    \# \text{ of } 3d \text{ modes} = \text{ceiling}(n/2).
\end{equation}
The modes \eqref{3d modes A class} correspond to BPS states that give rise to 3d $\mathcal{N}=2$ chiral multiplets. Therefore:
\begin{figure}[H]
    \centering
             \centering
    \scalebox{0.9}{\begin{tikzpicture}
    \node at (0,0) {M-theory on};
    \node at (0,-0.6) {$x^2+y^2+z^{n+1}+w^2+t^2=0$};
    \node at (4,-0.3) {$\Longleftrightarrow$};
    \node at (8.4,0) {3d $\mathcal{N}=2$ theory of};
    \node at  (8.4,-0.6) {ceiling$(n/2)$ uncharged chiral multiplets};
    %%%%%
    \end{tikzpicture}}\;
    \caption*{}
    \label{fig:summary A class}
    \end{figure}
    \vspace{-1.3cm}
As we have previously noticed, no flavor group for the 3d $\mathcal{N}=2$ theory is visible from the geometric perspective. This would substantiate the claim that the modes \eqref{3d modes A class} do not yield free 3d chiral multiplets. Nonetheless, unaccounted effects could enhance the rank of the flavor symmetry to make it compatible with the presence of free chirals, as we have previously mentioned. This puzzle has yet to be fully understood.\\

\indent Analogously to the cases presented in this section, M-theory reduced on the other members of the classes of $A_n$ singularities in Table \ref{ccDV table} engineers 3d $\mathcal{N}=2$ theories of uncharged chiral multiplets. As in our explicit examples, a simple recipe is followed in order to extract the number of chiral multiplets: an explicit $(n+1)\times(n+1)$ tachyon matrix is written down, and its fluctuations localized around the origin are computed. In order to classify the 3d theories associated with these singularities (similarly to what was done for the 5d SCFTs arising from threefolds in \cite{DeMarco:2022dgh}), the only non-trivial step one should face, that requires a slight external creative input, is finding the explicit form of the tachyon.
We leave this task for future work. 

In the next section, we tackle the ccDV isolated fourfolds constructed as deformed $D_n$ singularities.

\subsection{$D$ series}
Addressing M-theory geometric engineering on the singularities of type $D$ in Table~\ref{ccDV table} requires an additional tool, in order to allow the employment of tachyon condensation. As is very familiar, orientifolds are needed in Type IIA in order to produce singularities of type $D$ in the dual M-theory picture.\\
\indent For our purposes, we wish to describe D6-branes in the presence of orientifolds via the tachyon condensation technology, as outlined in \cite{Hori:2006ic,Diaconescu:2006id,Collinucci:2008pf}. Here, we follow the presentation that was laid down in \cite{Collinucci:2021wty} for the analogous cDV threefold context.\\

\indent The D6-branes live in the space parameterized by the complex variables $(\xi,w,t)$, and the orientifold action is produced by a $O6^-$ plane acting as:
\begin{equation}\label{O6 action}
    \sigma: \quad (\xi,w,t) \rightarrow (-\xi,w,t).
\end{equation}
The action \eqref{O6 action} naturally uplifts to the tachyon map between D8 and $\overline{\text{D8}}$-branes. The tachyon is defined as:
\begin{equation}
 \begin{tikzcd}
 F \arrow{r}{T} & E ,
\end{tikzcd}
\end{equation}
and its invariance under the orientifold action prescribes that
\begin{equation}\label{eq:orientifold inv tachyon}
    F \cong E^{\vee}, \quad T = -\sigma^*(T^t),
\end{equation}
with the $\vee$ superscript indicating the dual vector bundle. Compatibility with the presence of the orientifold dictates that the gauge redundancies on the D8/$\overline{\text{D8}}$-branes act on the tachyon as:
\begin{equation}\label{gauge orientifold}
    T \rightarrow G \cdot T \cdot \sigma^*(G^t).
\end{equation}
We are now in the right position to describe the uplift of the D6-brane locus to the full Calabi-Yau fourfold. We construct such fourfolds as $\mathbb{Z}_2$-quotients of a $\mathbb{C}^*$-fibration that degenerates on the brane locus. The $\mathbb{Z}_2$-quotient is required in order to take into account the orientifold projection, and it explicitly takes the form:
\begin{equation}\label{Z2 quotient}
    (\xi,w,t) \rightarrow (z:= \xi^2,w,t).
\end{equation}
The condition \eqref{eq:orientifold inv tachyon} implies that the tachyon can be uniquely written as:
\begin{equation}\label{eq:tachyon orientifold}
    T = A +\xi S, \quad \text{with } A^t = -A \quad \text{and} \quad S^t=S,
\end{equation}
as any term involving $\xi^2$ can be substituted thanks to \eqref{Z2 quotient}.
Furthermore, it is easy to ascertain that the double cover of the Calabi-Yau fourfold is written as:\footnote{The M-theory circle must belong to the $\mathbb{C}^*$ fiber $uv=c$. The orientifold must act as $u\leftrightarrow v$.}
\begin{equation}
    (x+i\xi y)(x-i\xi y) = P(\xi^2,w,t).
\end{equation}
Notice that the $\mathbb{C}^*$ acts on the two brackets in the left-hand side, while it acts trivially on the right-hand side. Taking the quotient we finally obtain the general form for our desired fourfold:\footnote{The most generic form of such a fibration is actually $x^2+zy^2 + y\,Q(z,w,t) + P(z,w,t)=0$ and the D6-brane locus, in this case, is $\Delta=Q^2+zP$.}
\begin{equation}\label{D fourfold}
    x^2+zy^2 + P(z,w,t)=0.
\end{equation}
Of course, the orientifold plane is located at $z=0$, and the D6-brane locus is where the quadric (in $y$) $zy^2+P(z,w,t)$ becomes a square, i.e. at
\begin{equation}\label{D brane locus}
    \Delta(z,w,t) = zP(z,w,t).
\end{equation}
The tachyon is a map that is non-invertible precisely on the D6-brane locus. Therefore:
\begin{equation}\label{spectral D}
    \text{det}(T) = \Delta(\xi^2,w,t) = \xi^2P(\xi^2,w,t).
\end{equation}

With the tachyon condensation toolbox at our disposal also in the presence of orientifolds, we can tackle the ccDV singularities of type $D$ in Table~\ref{ccDV table}. For the sake of conciseness, let us display the M-theory geometric engineering analysis for the class:
\begin{equation}\label{D class}
    x^2+zy^2+z^{n-1}+w^2+t^2=0.
\end{equation}
Analogously to the $A$ cases, in order to construct the tachyon we must first identify the $\mathbb{Z}_2$-covariant D6-brane locus. Comparing the general form of the fourfold \eqref{D fourfold} with our class \eqref{D class}, it is immediate to see that:
\begin{equation}
    \Delta(\xi^2,w,t) = \xi^2(\xi^{2(n-1)}+w^2+t^2).
\end{equation}
The task that is now at hand resides in correctly writing down the tachyon that produces the largest spectrum of states localized at $x = y = z = w = t =0$, that correspond to genuine 3d $\mathcal{N}=2$ modes.\\
\indent For the $D_n$ ccDV singularities \eqref{D class} the tachyon $T_n$ is an element of $O(2n,\mathbb{C})$. Besides, a quick close examination tells us that the tachyon yielding the maximal spectrum must also satisfy:
\begin{equation}
    T_n|_{\xi = w = t = 0} \quad \subset \quad [n,n],
\end{equation}
with $[n,n]$ the corresponding nilpotent orbit of $D_n$\footnote{If $n$ is even, we take the nilpotent orbit $[n,n]_I$, in the conventions of \cite{collingwood1993nilpotent}.}. E.g.\ in the $D_4$ case the tachyon explicitly takes the form\footnote{Notice that  $T$ takes the form $T=S+\xi A$. However, contrary to \eqref{eq:tachyon orientifold}, here we have chosen a basis such that the matrix $A$ is not antisymmetric but instead satisfies $A I + I A^T=0$ with
$I=\left(\begin{array}{c|c}
  \mathbb{0}_{n\times n}   & \mathbb{1}_{n\times n} \\
  \hline
  \mathbb{1}_{n\times n}   & \mathbb{0}_{n\times n} \\
\end{array}\right)$. In this presentation $S$ must satisfy $S I-I S^T=0$, that is fulfilled by $S=\mathbb{1}$~in~\eqref{eq:tachyonT4D4}.\label{bilinear form}}
\begin{equation}\label{eq:tachyonT4D4}
    T_4 = \left(
\begin{array}{cccc|cccc}
 \xi & 1 & 0 & 0 & 0 & 0 & 0 & 0 \\
 0 & \xi & 1 & 0 & 0 & 0 & 0 & 0 \\
 0 & 0 & \xi & 1 & 0 & 0 & 0 & \frac{(w+i T) }{2} \\
 0 & 0 & 0 & \xi & 0 & 0 & -\frac{(w+i T)}{2}  & 0 \\
 \hline
 0 & \frac{ (w-i T)}{2} & 0 & 0 & \xi & 0 & 0 & 0 \\
 -\frac{ (w- i T)}{2} & 0 & 0 & 0 & -1 & \xi & 0 & 0 \\
 0 & 0 & 0 & 0 & 0 & -1 & \xi & 0 \\
 0 & 0 & 0 & 0 & 0 & 0 & -1 & \xi \\
\end{array}
\right).
\end{equation}
Higher $D_n$ cases are constructed in a completely analogous fashion. Now we wish to identify the physical 3d $\mathcal{N}=2$ content arising from M-theory on the singularities described by the tachyons: to this end we compute fluctuations around the tachyons by linearizing \eqref{gauge orientifold}, i.e.:
\begin{equation}
    \delta T_n \sim \delta T_n + g\cdot T_n+T_n\cdot \sigma^* g^t,
\end{equation}
with $g \in \mathfrak{so}(2n,\mathbb{C})$.
Modding out explicitly by this equivalence relation we find that the tachyon $T_n$ localizes:
\begin{equation}
    \# \text{ of } 3d \text{ modes} = \text{floor}\left(\frac{n-1}{2}\right).
\end{equation}
As in the cases engineered by the $A$ series, the stabilizer of the tachyons $T_n$ is trivial, so that no flavor group arises in the 3d theory.\\
\indent Therefore, we can summarize the picture of geometric engineering on the singularities \eqref{D class} as:
\begin{figure}[H]
    \centering
             \centering
    \scalebox{0.9}{\begin{tikzpicture}
    \node at (0,0) {M-theory on};
    \node at (0,-0.6) {$x^2+zy^2+z^{n-1}+w^2+t^2=0$};
    \node at (4,-0.3) {$\Longleftrightarrow$};
    \node at (9,0) {3d $\mathcal{N}=2$ theory of};
    \node at  (9,-0.6) {floor$(\frac{n-1}{2})$ uncharged chiral multiplets};
    %%%%%
    \end{tikzpicture}}\;
    \caption*{}
    \label{fig:summary D class}
    \end{figure}
    \vspace{-1.3cm}
    Notice that the number of chirals does not match the rank of the flavor group that would be expected if they were free.\\
\indent All the other classes of type $D$ can be dealt with employing the exact same tools: the only small hurdle along the way lies in identifying the correct tachyon. We can now safely venture into the analysis of the exceptional cases.

\subsection{Exceptional cases and the adjoint scalar approach}\label{eq:isolExcepCases}
The investigation of the ccDV singularities of type $E_6,E_7,E_8$ in Table~\ref{ccDV table} presents additional hurdles that are more of a technical nature, rather than a conceptual one. The main issue is that no $\mathbb{C}^*$-fibration can be identified in these singularities, thus preventing the analysis from the tachyon condensation perspective. This is dual to the statement that M-theory on a non-compact K3 with $E_6,E_7,E_8$ singularities, namely 7d SYM with $G=E_6,E_7,E_8$, does not admit a description in terms of D6-branes. The way out of this conundrum was already identified in \cite{Collinucci:2022rii,DeMarco:2022dgh} with eyes set specifically on the cDV threefold cases. Thankfully, nothing stands in the way of generalization of that method to the fourfold case. Let us quickly review it, before examining an explicit exceptional fourfold singularity.\\
\indent For the ccDV singularities constructed as deformations of $A$ and $D$ singularities, we have described the corresponding D6-brane configurations via the tachyon map. The relation between the tachyon and the brane-locus can be identified thanks to the $\mathbb{C}^*$-fibration enjoyed by such singularities, and is made explicit through the spectral equations of the form \eqref{D6 quadrifold} and \eqref{spectral D}. One could have reformulated this relation in the following terms: M-theory on K3 with $A$ and $D$ Du Val singularities yields 7d SYM with gauge group of type $A$ and $D$ respectively. In this theory, the vector multiplet hosts an adjoint scalar $\Phi$, whose vev describes deformations of the D6-brane stack from the type IIA perspective. The adjoint scalar $\Phi$ is of course closely related to the tachyon: following \cite{Collinucci:2014qfa}, one can always rewrite the tachyon as
\begin{equation}\label{tachyon higgs}
\begin{split}
   & A\text{ case:} \quad T(z,w,t) = z\mathbb{1}-\Phi(w,t),\\
   & D\text{ case:} \quad T(\xi,w,t) = \xi\mathbb{1}-\Phi(w,t),
    \end{split}
\end{equation}
for some choice of the matrices $\Phi(w,t)$. Therefore, the spectral equations for the $A$ and $D$ cases can be rewritten in full generality as:
\begin{table}[H]
\scalemath{0.9}{
\begin{array}{c|c|c|c}
   \text{Singularity}  & \text{Tachyon} & \text{Adjoint scalar }\Phi(w,t) & \text{ccDV equation}\\
   \hline
  A & \Delta = \text{det}(T(z,w,t)) & \Delta = \text{det}(z\mathbb{1}-\Phi(w,t)) & x^2+y^2+\Delta(z,w,t) = 0\\
  \hline
  D & \Delta = \text{det}(T(\xi,w,t)) & \Delta = \text{det}(\xi\mathbb{1}-\Phi(w,t)) & x^2+zy^2 +\frac{\Delta-\text{Pfaff}^2(\Phi)}{z}+2y\text{Pfaff}(\Phi) = 0 \\
\end{array}}
\caption{Spectral equation for deformed $A$ and $D$ singularities.}
\label{table spectral}
\end{table}
As we have mentioned, for the exceptional cases $E_6,E_7,E_8$ there is no Type IIA dual in terms of D6-branes. Nonetheless, M-theory on such surfaces leads to 7d SYM with gauge group of type $E_6,E_7,E_8$. This also has an adjoint scalar $\Phi$. Switching on a vev with non-trivial profile for $\Phi$ corresponds to a deformation of the $E_6,E_7,E_8$ singularity.
The Casimirs of $ \Phi$ can be rigorously related to deformations of the $E_6,E_7,E_8$ Du Val singularities, thus producing ccDV singularities. Namely we could add the three rows appearing in Table~\ref{E table spectral} to Table~\ref{table spectral}:
\begin{table}[H]
\scalemath{0.9}{
\begin{array}{c|c|c|c}
   \text{Singularity}  & \text{Tachyon} & \text{Adjoint scalar }\Phi(w,t) & \text{ccDV equation}\\
   \hline
  E_6 & \emptyset & \makecell{\mu_2(\Phi),\mu_5(\Phi),\mu_6(\Phi),\\
  \mu_8(\Phi),\mu_9(\Phi),\mu_{12}(\Phi)} & \makecell{x^2+z^4+y^3+\mu_2 y z^2+\mu_5 y z+\\
  +\mu_6 z^2+\mu_8 y+\mu_9 z+\mu_{12}=0}\\
  \hline
  E_7 & \emptyset & \makecell{\tilde{\mu}_2(\Phi),\tilde{\mu}_6(\Phi),\tilde{\mu}_8(\Phi),\tilde{\mu}_{10}(\Phi)\\
  \tilde{\mu}_{12}(\Phi),\tilde{\mu}_{14}(\Phi),\tilde{\mu}_{18}(\Phi)} & \makecell{x^2+y^3+y z^3+\tilde{\mu}_2 y^2 z+\tilde{\mu}_6 y^2+\tilde{\mu}_8 y z+\\
+ \tilde{\mu}_{10} z^2+\tilde{\mu}_{12} y+\tilde{\mu}_{14} z+\tilde{\mu}_{18}=0}\\
  \hline  E_8 & \emptyset & \makecell{\hat{\mu}_2(\Phi),\hat{\mu}_8(\Phi),\hat{\mu}_{12}(\Phi),\hat{\mu}_{14}(\Phi)\\
  \hat{\mu}_{18}(\Phi),\hat{\mu}_{20}(\Phi),\hat{\mu}_{24}(\Phi),\hat{\mu}_{30}(\Phi)} & \makecell{x^2+y^3+z^5+\hat{\mu}_2 y z^3+\hat{\mu}_8 y z^2+\hat{\mu}_{12} z^3+\\
  +\hat{\mu}_{14} y z+\hat{\mu}_{18} z^2+\hat{\mu}_{20} y+\hat{\mu}_{24} z+\hat{\mu}_{30}=0}\\
\end{array}}
\caption{Spectral equation for deformed exceptional singularities.}
\label{E table spectral}
\end{table}
For the explicit dependence of the deformation parameters $\mu_i,\tilde{\mu}_i,\hat{\mu}_i$ in terms of the Casimirs of $\Phi$, we refer to Appendix C of \cite{DeMarco:2022dgh}.

The localized modes can be computed as fluctuations of $\Phi$ modulo linearized gauge transformations, i.e.\
$\varphi \in \mathfrak{g}$ such that (see \cite{Collinucci:2021ofd,Collinucci:2022rii})
\begin{equation}
\partial \varphi = 0, \quad \varphi \sim \varphi + [\Phi , g]   \:, 
\end{equation}
with $\mathfrak{g}$ the Lie algebra where $\Phi$ lives in and $g$ a generic element in $\mathfrak{g}$. 
As it can be understood from the relation between $T$ and $\Phi$ (see \eqref{tachyon higgs}) the fluctuations of $\Phi$ are in one-to-one correspondence to those of $T$ \cite{Collinucci:2014qfa}.\\

\indent We now finally possess all the tools needed to analyze an explicit example of exceptional ccDV singularity. Let us pick the following fourfold singularity from Table~\ref{ccDV table}:
\begin{equation}\label{E6 case}
    x^2+y^3+z^4+w^2+t^2 = 0.
\end{equation}
The adjoint scalar field $\Phi$ that reproduces \eqref{E6 case} via the generalized spectral equation \eqref{E table spectral} can be written in terms of the generators $e_{\alpha_i}$ of the simple roots $\alpha_i$ of $E_6$ (see Figure~\ref{fig:dynkinE6} for our choice of convention):
%\footnote{We adopt as convention the following notation for the simple roots: 
\begin{figure}[H]
    \centering
$\scalebox{0.7}{\begin{tikzpicture}
        %\draw[thick] (0,0) circle (0.65);
        \node at (0,0) {\Large$\boldsymbol{\alpha_3}$};
        \draw[thick] (0.7,0)--(1.3,0);
        %\draw[thick] (2,0) circle (0.65);
        \node at (2,0) {\Large$\boldsymbol{\alpha_4}$};
        \draw[thick] (-0.7,0)--(-1.3,0);
        %\draw[thick] (-2,0) circle (0.65);
        \node at (-2,0) {\Large$\boldsymbol{\alpha_2}$};
        \draw[thick] (0,0.7)--(0,1.3);
        %\draw[thick] (0,2) circle (0.65);
        \node at (0,2) {\Large$\boldsymbol{\alpha_6}$};
        \draw[thick] (2.7,0)--(3.45,0);
        %\draw[thick] (3.5,-0.5)--(4.5,-0.5)--(4.5,0.5)--(3.5,0.5)--cycle ;
        \node at (4,0) {\Large$\boldsymbol{\alpha_5}$};
        \draw[thick] (-2.7,0)--(-3.45,0);
        %\draw[thick] (-3.5,-0.5)--(-4.5,-0.5)--(-4.5,0.5)--(-3.5,0.5)--cycle ;
        \node at (-4,0) {\Large$\boldsymbol{\alpha_1}$};
        %\draw[thick] (0,2.7)--(0,3.45);
        %\draw[thick] (-4.7,0)--(-5.3,0);
        %\draw[thick] (4.7,0)--(5.3,0);
        %\draw[thick] (0,-0.7)--(0,-1.3);
        \end{tikzpicture}}$
            \caption{Dynkin diagram of $E_6$.}
    \label{fig:dynkinE6}
\end{figure}
\begin{equation}
    \Phi(w,t) = (iw+t)e_{\alpha_1}+ie_{\alpha_2}+e_{\alpha_3}+e_{\alpha_4}+e_{\alpha_5}+e_{\alpha_6}+(iw-t)e_{-\alpha_1-2\alpha_2-3\alpha_3-2\alpha_4-\alpha_5-2\alpha_6} .
\end{equation}
Its zero modes can be computed by using the code developed in \cite{Collinucci:2022rii}. As a result, one finds 
2 modes localized at $x=y=z=w=t=0$, i.e.
\begin{figure}[H]
    \centering
             \centering
    \scalebox{0.9}{\begin{tikzpicture}
    \node at (0,0) {M-theory on};
    \node at (0,-0.6) {$x^2+y^3+z^4+w^2+t^2=0$};
    \node at (4,-0.3) {$\Longleftrightarrow$};
    \node at (8,0) {3d $\mathcal{N}=2$ theory of};
    \node at  (8,-0.6) {2 uncharged chiral multiplets};
    %%%%%
    \end{tikzpicture}}\;
    \caption*{}
    \label{fig:summary E6 class}
    \end{figure}
    \vspace{-1.3cm}

\subsection{Zero modes and the deformed phase}\label{sec: deformations}

In the preceding sections we have shown that the classes of isolated ccDV singularities listed in Table \ref{ccDV table} admit a description through a tachyon field $T$ or, equivalently, by means of an adjoint scalar field $\Phi$, as summarized in Table \ref{table spectral} and \ref{E table spectral}. Computing fluctuations of $\Phi$ we have identified the zero-modes that are localized on the singular point. These correspond to chiral multiplets in the 3d $\mathcal{N}=2$ theory engineered by M-theory on the singular fourfold.\\
\indent It is natural to switch on a vev for said chiral multiplets and detect its effects on the fourfold geometry. We momentarily show that these vev's are related to deformations of the singular fourfold, and extract the physical meaning of this observation.\\
\indent To ground these ideas, consider the example introduced in \eqref{quadrifold}, that we rewrite here for convenience, along with the adjoint scalar field $\Phi$ that describes it (it can be obtained directly from the tachyon \eqref{tachyon quadrifold} by removing the part proportional to the identity, thanks to \eqref{tachyon higgs}). For the ccDV $A$ cases, the singular equation is recovered through the spectral equation from Table \ref{table spectral}:
\begin{equation}
    x^2+y^2+\text{det}(z\mathbb{1}-\Phi) =0.
\end{equation}
Therefore, for the fourfold \eqref{quadrifold} the singular equation and the corresponding $\Phi$ read:
\begin{equation}\label{singular fourfold}
    X: \quad x^2+y^2+z^2+w^2+t^2 = 0 \quad \longleftrightarrow \quad \Phi =\left(\begin{array}{cc}
     i w   & t \\
       -t  & -iw
    \end{array} \right) \:.
\end{equation}
Now consider switching on a vev $\boldsymbol{a}$ for the only zero-mode that is localized on $x=y=z=w=t=0$, accordingly modifying $\Phi$. This produces a deformation of the singular fourfold \eqref{singular fourfold}, as can be seen explicitly employing the spectral equation:
\begin{equation}\label{deformed fourfold}
    X: \quad x^2+y^2+z^2+w^2+t^2= \boldsymbol{a^2}  \quad \longleftrightarrow \quad \Phi =\left(\begin{array}{cc}
     i w   & t+\boldsymbol{a} \\
       -t+\boldsymbol{a}  & iw
    \end{array} \right) \:.
\end{equation}
The deformed equation in \eqref{deformed fourfold} is precisely the versal deformation of the singular fourfold \eqref{singular fourfold}, as can be easily checked computing the Milnor number\footnote{The Milnor number is defined as the complex dimension of the coordinate ring $\mathbb{C}[x,y,z,w,t]/(\frac{\partial f}{\partial x},\frac{\partial f}{\partial y},\frac{\partial f}{\partial z},\frac{\partial f}{\partial w},\frac{\partial f}{\partial t})$, with $f$ the hypersurface equation defining the singularity.} $\mu = 1$.\\
\indent Things start growing more involved in the next-to-simplest example, examined in Section~\ref{sec: A series}:
\begin{equation}\label{singular fourfold 2}
    X: \quad x^2+y^2+z^3+w^2+t^2 = 0 \quad \longleftrightarrow \quad \Phi =\left(\begin{array}{ccc}
      0   & 1 & 0 \\
       0 & 0 & w+it \\
       w-it & 0 & 0 \\
    \end{array} \right) \:.
\end{equation}
The corresponding 3d theory possesses only one chiral multiplet, and turning on a vev for it produces the  following deformed singularity:
\begin{equation}\label{deformed fourfold 2}
    X: \quad x^2+y^2+z^3+w^2+t^2 = \boldsymbol{z a^2 }\quad \longleftrightarrow \quad \Phi =\left(\begin{array}{ccc}
      0   & 1 & 0 \\
       0 & \boldsymbol{a} & w+it \\
       w-it & 0 & -\boldsymbol{a} \\
    \end{array} \right) \:.
\end{equation}
In this case, the deformed fourfold \eqref{deformed fourfold 2} is not the full versal deformation of \eqref{singular fourfold 2}, which would instead read (given that $\mu = 2$ for the singularity at hand):
\begin{equation}
    x^2+y^2+z^3+w^2+t^2 = z \lambda_1+\lambda_2.
\end{equation}
The mismatch between deformations parameters and zero-modes appears in a similar fashion in all the isolated ccDV singularities examined so far. 

In the M-theory on CY3 context, it was possible to match deformations of cDV singularities with the number of hypermultiplets of the corresponding 5d SCFTs, carefully tracking paired and unpaired deformations \cite{Closset:2020scj,Collinucci:2021ofd}.
In the fourfold context the picture is murkier: it is crucial to remark that the work of \cite{Gukov:1999ya}\footnote{See also the related seminal work \cite{Shapere:1999xr}.} proved that the ccDV geometries such as \eqref{singular fourfold} and \eqref{singular fourfold 2} \textit{do not admit normalizable deformations}. Let us quickly recap their argument here.\\
\indent Consider a quasi-homogeneous ccDV singularity with $\mathbb{C}^*$ action and weights as in \eqref{C star weights}, defined by the hypersurface equation $f(x,y,z,w,t) = 0$. Then its versal deformation takes the form:
\begin{equation}
    f(x,y,z,w,t) + \sum_i \lambda_i g_i(x,y,z,w,t)=0,
\end{equation}
for a well-defined choice of the polynomials $g_i$,  and with the $\lambda_i$ complex parameters characterizing the deformation ($i=1,\ldots,\mu$, where $\mu$ is the Milnor number of the singularity). Call $s_i$ the scaling dimension of the $g_i$ under the $\mathbb{C}^*$ action. Then, the deformation parameterized by $\lambda_i$ is \textit{normalizable} if and only if:
\begin{equation}
    (\omega_x+\omega_y+\omega_z+\omega_w+\omega_t)+s_i \leq 2.
\end{equation}
It is easy to check that \eqref{singular fourfold} and \eqref{singular fourfold 2} do not admit normalizable deformations. Other singularities from Table \ref{ccDV table}, though admit one or more normalizable deformations. Nonetheless, the number of localized modes that are found via our techniques is always equal or greater than the number of normalizable deformations à la \cite{Gukov:1999ya}. One such example is the fourfold singularity
\begin{equation}
    x^2+y^2+z^5+w^3+t^3 = 0,
\end{equation}
that admits 2 normalizable deformations and 3 localized modes, as can be readily seen constructing an explicit tachyon and computing its fluctuations with the recipe of Section~\ref{sec: A series}.\\
\indent From the physical perspective, this produces a puzzle: the zero-modes in the singular fourfold phase are normalizable, as they are localized at the origin, and possess a well-defined finite kinetic term. On the other hand, when a non-trivial vev for the zero-modes is switched on, they produce a deformation of the geometry. Since such deformations are non-normalizable due to \cite{Gukov:1999ya}, the fluctuations around the vev become non-dynamical and thus do not give rise to a full-fledged branch of the moduli space of the 3d theory, but rather to a set of couplings. This puzzle has already been noted in the recent work \cite{Acharya:2024bnt} in the context of 5d theories of hypermultiplets, and it is clear that it requires further attention in future research.

%{\color{purple} RV: qui dobbiamo dire che questi CY possono essere smoothati deformando, parlare un po' delle deformazioni, mostrare che dando vev ai modi che abbiamo trovato sul tachione il CY4 viene deformato (riportare esempio semplice), risultato di GVW che per questi CY4 deformati, i modi di sugra relativi ai c.s. deformation non  sono normalizzabili cio\`e hanno termine cinetico che esplode e qndi sono coupling, concludiamo che quindi nel limite singolare il termine cinetico deve rimanere finito e che esplode appena accendiamo deformazione citando paper bobby.}

\section{3d $\mathcal{N}=2$ theories from non-isolated Calabi-Yau fourfolds}\label{sec: non iso 3d SCFTs}
We now venture into the analysis of the class of non-isolated fourfold terminal singularities introduced in Section~\ref{sec: non iso curve singularities}. From the physical point of view, these retain all the advantages of the ccDV singularities: given that no compact 4- or 6-cycle is admitted in any of their crepant resolutions, as shown in Section~\ref{sec: non iso curve singularities} by performing an explicit blow-up, no instanton corrections and contributions of the $G_4$ flux are present. At the same time, the non-isolated singularities at hand overcome the limitation of the isolated ccDV class: as we will explicitly see, the corresponding 3d $\mathcal{N}=2$ theory generally features a flavor symmetry of type $U(1)^{\ell}$, with a non-empty spectrum of states charged under it.

\subsection{A series}\label{sec: A series v2}

\begin{figure}[t!]
    \centering
$\scalebox{0.6}{\begin{tikzpicture}
        \draw[thick, fill = black] (0,0) circle (0.4);
        \node at (0,-0.65) {$n$-th node};
        \draw[thick] (0.7,0)--(1.3,0);
        %\draw[thick] (2,0) circle (0.4);
        %\node at (2,0) {\Large$\boldsymbol{\alpha_4}$};
        \draw[thick] (-0.7,0)--(-1.3,0);
        \node at (-2,0) {$\cdots$};
        %\draw[thick] (-2,0) circle (0.4);
        %\node at (-2,0) {\Large$\boldsymbol{\alpha_2}$};
        \draw[thick] (4,0) circle (0.4);
        \draw[thick] (-4,0) circle (0.4);
        \node at (2,0) {$\cdots$};
        \draw[thick] (2.7,0)--(3.45,0);
        %\draw[thick] (3.5,-0.5)--(4.5,-0.5)--(4.5,0.5)--(3.5,0.5)--cycle ;
      %  \node at (4,0) {\Large$\boldsymbol{\alpha_5}$};
        \draw[thick] (-2.7,0)--(-3.45,0);
        %\draw[thick] (-3.5,-0.5)--(-4.5,-0.5)--(-4.5,0.5)--(-3.5,0.5)--cycle ;
        %\node at (-4,0) {\Large$\boldsymbol{\alpha_1}$};
        %\draw[thick] (0,2.7)--(0,3.45);
        %\draw[thick] (-4.7,0)--(-5.3,0);
        %\draw[thick] (4.7,0)--(5.3,0);
        %\draw[thick] (0,-0.7)--(0,-1.3);
        \end{tikzpicture}}$
            \caption{Dynkin diagram of $A_{2n-1}$, with the resolved node highlighted in black.}
    \label{fig:dynkinA}
\end{figure}
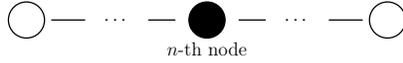

\indent In order to make our arguments concrete, we focus on the following class of non-isolated singularities, obtained from a base-change of a cDV singularity of type $A$:
\begin{equation}\label{4fold pagoda}
    x^2+y^2+z^{2n}-w^2t^2 = 0 \quad \subset \quad \mathbb{C}^5,
\end{equation}
which is singular at
\begin{equation}\label{A sing lines}
 x = y = z = w = 0,\quad\quad x = y = z = t = 0,
\end{equation}
with an enhancement of the singularity at the intersection point. We will discuss this aspect in more details momentarily.\\
\indent Concretely tackling the analysis of the spectrum of M-theory on \eqref{4fold pagoda} entails identifying the corresponding tachyon. We can easily highlight the D6-brane locus writing \eqref{4fold pagoda} as:
\begin{equation}
    uv = (z^n-wt)(z^n+wt).
\end{equation}
Thus:
\begin{equation}\label{k brane locus}
    \Delta_n = (z^n-wt)(z^n+wt).
\end{equation}
Notice that the brane locus factorizes: this is the telltale sign that a partial resolution inflating a $\mathbb{P}^1$ is possible, as was shown in Section~\eqref{sec: non iso curve singularities}; physically, it implies that a non-trivial flavor symmetry survives on the D6-brane stack, even when the tachyon is switched on. We will prove this latter statement momentarily.\\
\indent Following the teachings of the previous sections, it is straightforward to find the tachyon describing the D6-brane configuration \eqref{k brane locus}:
\begin{equation}\label{non iso tachyon}
    T_n = \scalemath{0.8}{\left( \begin{array}{ccccccccc|ccccccccc}
      z & 1 & 0 & &\cdots & & & & 0 & & & & & & & & &\\
     0 & z & 1 & 0 & \cdots & & & & 0 & & & & & & & & &\\
       &  &  & \ddots & & & & & & & & & & & & & &\\
    \vdots & \vdots & & z & w & 0 & & & \vdots & & & & & \mathbb{0}_{n\times n} & & & &\\
    & &  & & z & 1 & 0 & & & & & & & & & & & \\
        &  &  &  &  &  & \ddots & & & & & & & & & & &\\
      0 & & & &\cdots  & & z & & 1 & & & & & & & & &\\
       t & 0 &   & & \cdots& & & & z & & & & & & & & &\\
    \hline
 &  & & & & & & & &   z & 1 & 0 & &\cdots & & & & 0 \\
 & & & & & & & & &   0 & z & 1 & 0 & \cdots & & & & 0\\
 & & & & & & & & &     &  &  & \ddots & & & & & \\
 & & &\mathbb{0}_{n\times n} & & & & & &   \vdots & \vdots & & z & w & 0 & & & \vdots \\
 & & & & & & & & &   & &  & & z & 1 & 0 & &\\
 & & & & & & & & &       &  &  &  &  &  & \ddots & &\\
 & & & & & & & & &     0 & & & &\cdots  & & z & & 1\\
& & & & & & & & &      -t & 0 &   & & \cdots& & & & z\\
    \end{array} \right)}
\end{equation}
We wish to produce the 3d $\mathcal{N}=2$ theory with the maximal spectrum: to this end, we take the tachyons $T_n$ to be related to a specific nilpotent orbit:
\begin{equation}
    T_n|_{z=w=t=0} \quad \subset \quad \left[\text{ceiling}\left(\frac{n}{2}\right),\text{ceiling}\left(\frac{n}{2}\right),\text{floor}\left(\frac{n}{2}\right),\text{floor}\left(\frac{n}{2}\right)\right]. \nonumber
\end{equation}
One can understand the rationale behind this choice noticing that this is the nilpotent orbit of maximal codimension compatible with our brane locus. In turn, this implies that the maximal spectrum in 3d is obtained. For further details on this connection, we refer to \cite{DeMarco:2021try}. \\
\indent A relevant observation is in order: the tachyon \eqref{non iso tachyon} is composed of two $n\times n$ blocks, each identical to the tachyons that we have introduced for isolated ccDV singularities in \eqref{iso tachyon}. This is a natural consequence of our discussion on the crepant resolution of our class of the non-isolated singularities \eqref{4fold pagoda} that we have exhibited in Section~\ref{sec: non iso curve singularities}. Indeed, as we have shown in that Section, the fourfold admits a partial resolution that inflates a $\mathbb{P}^1$ on top of each point of the singular lines \eqref{A sing lines}, corresponding to the central node of the $A_{2n-1}$ Dynkin diagram, visible in Figure \ref{fig:dynkinA}.
This can be foreseen, in alternative to an explicit blowup computation à la Section~\ref{sec: non iso curve singularities}, noticing that the part of the tachyon \eqref{non iso tachyon} that is not proportional to the identity (this is the adjoint scalar field $\Phi$ we have introduced in Section~\ref{eq:isolExcepCases}, see \eqref{tachyon higgs}) belongs to a subalgebra  $A_{n-1}\oplus A_{n-1} \subset A_{2n-1}$ (for a more in-depth 
analysis, we refer the analogous case of threefolds examined in \cite{DeMarco:2022dgh}).\\
\indent The partially resolved fourfold still displays two isolated singular points lying on the north and south pole of the fiber on top of the origin, where the singularity enhances, and that cannot be resolved further. These are precisely two ccDV points of the form
\begin{equation}\label{extra points}
    x_1^2+x_2^2+x_3^n+x_4^2+x_5^2=0,
\end{equation}
each of which is engineered by a $n\times n$ tachyon of the form \eqref{iso tachyon}. We will see that this fact influences the 3d $\mathcal{N}=2$ content in a predictable fashion.\\

\indent We now look for fluctuations of the tachyon \eqref{non iso tachyon} that localize at the origin $u=v=z=w=t=0$: these correspond to the genuine physical content of the 3d $\mathcal{N}=2$ theories engineered by the very same tachyon.
As we have amply seen in previous sections, this is achieved by removing the gauge redundancies inherited from the D8-$\overline{\text{D8}}$ stack. The generic fluctuations can be organized as:
\begin{equation}\label{k fluctuations}
    \delta T_n = \left(\begin{array}{cc}
      ( \delta t_{11})_{n\times n}  &( \delta t_{12})_{n\times n}   \\
     ( \delta t_{21})_{n\times n}     & ( \delta t_{22})_{n\times n} 
    \end{array} \right),
\end{equation}
where the $\delta t_{ij}$ are $n\times n$ matrices.
Let us also notice that the stabilizer of $T_n$ is non-trivial, namely the tachyon preserves a subgroup of the symmetry on the D8-$\overline{\text{D8}}$ stack:
\begin{equation}
    \text{Stab}(T_n) = U(1).
\end{equation}
This $U(1)$ factor is a combination of the two $U(1)$'s living on the two non-compact D6-branes and 
is interpreted as a flavor group from the point of view of the 3d theory sitting at the origin of the brane locus \eqref{k brane locus}, and it is embedded in $GL(2n,\mathbb{C})$ as (employing the same representation we used for $T_n$):
\begin{equation}\label{U1 embedding}
    G_{\text{flavor}} = \left(\begin{array}{cc}
       g\mathbb{1}_{n\times n}  &\mathbb{0}_{n\times n}   \\
     \mathbb{0}_{n\times n}     & g^{-1}\mathbb{1}_{n\times n} 
    \end{array} \right)
\end{equation}
with $g$ a parameter in $\mathbb{C}$.
We can now proceed to detect the 3d physical content. It is a completely algorithmic process that yields the following results:
\begin{eqnarray}
&&  \# \text{ of } 3d \textit{ uncharged}\text{ modes} = 2\cdot\text{ceiling}\left(\frac{n-1}{2}\right), \nonumber \\
&& \#   \text{ of } 3d \textit{ charged}\text{ modes} = 2\cdot\text{ceiling}\left(\frac{n-1}{2}\right).  \nonumber
\end{eqnarray}
\textit{Uncharged} modes correspond to fluctuations in $( \delta t_{11})_{n\times n}$ and $( \delta t_{22})_{n\times n}$ of \eqref{k fluctuations}. These can be interpreted exactly in the same fashion as the 3d uncharged modes we highlighted in the ccDV singularity case of Section~\ref{sec: 3d SCFTs}: they sit on top of the singular isolated points \eqref{extra points} that are leftover after the partial resolution of our non-isolated singularities. Their appearance could have alternatively been forecast by the fact that the two blocks of the tachyons $T_n$ in \eqref{non iso tachyon} are identical, modulo a trivial change of variables, to the tachyons of ccDV singularities \eqref{iso tachyon}.\\
\indent On the other hand, \textit{charged} modes sit in the fluctuations $( \delta t_{12})_{n\times n}$ and $( \delta t_{21})_{n\times n}$ of \eqref{k fluctuations}. It is easy to see that the $U(1)$ flavor group \eqref{U1 embedding} acts non-trivially on such fluctuations: therefore, the charged zero-modes have charge $\pm 1$ respectively (up to an overall normalization factor) under it.\\
\indent All in all, the 3d $\mathcal{N}=2$ theory engineered by our class of non-isolated fourfold singularities reads:
\begin{figure}[H]
    \centering
             \centering
    \scalebox{0.9}{\begin{tikzpicture}
    \node at (0,0) {M-theory on};
    \node at (0,-0.6) {$x^2+y^2+z^{2n}-w^2t^2=0$};
    \node at (3.2,-0.3) {$\Longleftrightarrow$};
    \node at (8,0) {3d $\mathcal{N}=2$ theory with $G_{\textit{flavor}}=U(1)$ of};
    \node at  (8,-0.6) {\text{2$\cdot$ceiling}$\left(\frac{n-1}{2}\right)$ uncharged chiral multiplets};
    \node at  (8.2,-1.2) {\text{and 2$\cdot$ceiling}$\left(\frac{n-1}{2}\right)$ charged chiral multiplets};
    %%%%%
    \end{tikzpicture}}\;
    \caption*{}
    \label{fig:summary non iso class}
    \end{figure}
    \vspace{-1.3cm}
An important remark naturally arises given the presence of matter that is charged under the flavor symmetry in the 3d $\mathcal{N}=2$ theory. Let us briefly recall how things work in the threefold case: M-theory on terminal cDV singularities engineers rank 0 theories composed of a bunch of hypermultiplets. If the threefold admits a complete resolution (i.e.\ no singularity is left in the threefold after the resolution) the number of charged hypermultiplets is in one-to-one correspondence with the genus zero Gopakumar-Vafa (GV) invariants of the threefolds.\footnote{For works exploring this connection for rank 0 5d SCFTs, see \cite{Collinucci:2021wty,Collinucci:2022rii}. See \cite{Gopakumar:1998ii,Gopakumar:1998ki,Gopakumar:1998jq} for the original works on GV invariants in the CY3 case, and \cite{Klemm:2007in} for the seminal work on the CY4 setting.} I.e., in the simple case where only one $U(1)$ flavor factor is present, namely the resolution has produced a single $\mathbb{P}^1$ as exceptional set (the generalization to multiple factors is effortless):
\begin{equation}
    \# \text{ of hypers of charge }d \quad \longleftrightarrow \quad n^0_d(\text{CY3}),
\end{equation}
with $n^0_d(\text{CY3})$ the genus zero GV invariants of the cDV threefold for the class $d[\mathbb{P}^1]$. It is hence natural to ponder whether a similar relationship might hold in the fourfold cases we have examined so far. Such a conjectural relation would connect the number of charged chiral multiplets in the 3d $\mathcal{N}=2$ theory and the GV invariants of the corresponding non-isolated CY4, yielding something of the sort:
\begin{equation}
     \# \text{ of chirals of charge }d \quad \xleftrightarrow{\hspace{0.3cm}\textbf{?}\hspace{0.3cm}} \quad n^0_d(\text{CY4}).
\end{equation}
 A clear difficulty in testing this conjecture resides in the explicit computation of the GV invariants from the CY4 perspective, which is particularly hard given the non-toric setting we are operating in, and that in general is not even sensible if the resolved model is not completely smooth. To this end, the recent mathematical literature developed in \cite{Cao:2014bca,Nekrasov:2017cih,Nekrasov:2018xsb,Cao:2018wmd,Kononov:2019fni,Cao:2019fqq,Cao:2019tnw,Cao:2020vce,Cao:2020otr,Cao:2020hoy,Bousseau:2020fus,cao2022k,monavari2022canonical,cao2023donaldson,Piazzalunga:2023qik,Nekrasov:2023nai,Liu:2024bgp} could provide a solid starting point.\\

\indent Finally, let us notice that the analysis of M-theory geometric engineering on more general classes of fourfold singularities, constructed from a base-change of cDV singularities, can be tackled in a similar fashion. In general, they take the form:
\begin{equation}\label{general case 2}
        x^2+y^2+z^{n}+w^{k_1}t^{k_2}= 0
    \quad \text{or } \quad 
        x^2+y^2+z(z^{n-1}+ w^{k_1}t^{k_2})= 0,
\end{equation}
which have two non-compact lines of singularities (for $n,k_1,k_2 >1$) at
\begin{equation}\label{non compact lines}
    x=y=z=w=0, \quad x=y=z=t=0,
\end{equation}
with enhancement of the singularity at the intersection point.\\
\indent The corresponding tachyon will be a $n\times n$ matrix. For generic values of $n,k_1,k_2$, the stabilizer of the tachyon describing the singularity on the left-hand side of \eqref{general case 2} is trivial, so that no flavor group arises from the 3d $\mathcal{N}=2$ perspective, and is simply $U(1)$ in the case on the right-hand side. For specific choices of $n,k_1,k_2$, instead, the tachyon factorizes in $\ell+1$ pieces. For such cases one finds that it can be written in block diagonal form:
\begin{equation}
    T = \left(\begin{array}{cccc}
     \mathcal{B}_1    & & & \\
         & \mathcal{B}_2 & & \\
          & & \ddots & \\
        & & & \mathcal{B}_{\ell+1}\\
    \end{array} \right)\:.
\end{equation}
Each factor corresponds to a single stack of D6-branes that have recombined into a single brane as a consequence of the tachyon vev: hence each of the blocks $\mathcal{B}_i$ engineers on its own a singularity of type $A$ from Table~\ref{ccDV table}. These singularities are leftover after the partial resolution (if any) of the non-compact lines \eqref{non compact lines}. In such cases, the stabilizer of the tachyon is $U(1)^{\ell}$. Therefore, the 3d theory displays
\begin{equation}
    G_{\text{flavor}} = U(1)^{\ell},
\end{equation}
as well as charged and uncharged hypers. The charges can be readily computed via the explicit embedding of $U(1)^{\ell}$ in $GL(n,\mathbb{C})$. In the preceding example engineered by \eqref{4fold pagoda}, we have exhibited a class with $\ell = 1$.

\subsection{$D$ series}\label{sec: D series}
\begin{figure}[H]
    \centering
$\scalebox{0.6}{\begin{tikzpicture}
        \draw[thick] (0,0) circle (0.4);
        %\node at (0,0) {\Large$\boldsymbol{\alpha_3}$};
        \draw[thick] (0.7,0)--(1.3,0);
        \draw[thick, fill=black] (2,0) circle (0.4);
        %\node at (2,0) {\Large$\boldsymbol{\alpha_4}$};
        \draw[thick] (-0.7,0)--(-1.3,0);
        \draw[thick, fill = black] (-2,0) circle (0.4);
        %\node at (-2,0) {\Large$\boldsymbol{\alpha_2}$};
        \draw[thick] (0,0.7)--(0,1.3);
        \draw[thick] (0,2) circle (0.4);
        %\node at (0,2) {\Large$\boldsymbol{\alpha_6}$};
        %\draw[thick] (2.7,0)--(3.45,0);
        %\draw[thick] (3.5,-0.5)--(4.5,-0.5)--(4.5,0.5)--(3.5,0.5)--cycle ;
      %  \node at (4,0) {\Large$\boldsymbol{\alpha_5}$};
        %\draw[thick] (-2.7,0)--(-3.45,0);
        %\draw[thick] (-3.5,-0.5)--(-4.5,-0.5)--(-4.5,0.5)--(-3.5,0.5)--cycle ;
        %\node at (-4,0) {\Large$\boldsymbol{\alpha_1}$};
        %\draw[thick] (0,2.7)--(0,3.45);
        %\draw[thick] (-4.7,0)--(-5.3,0);
        %\draw[thick] (4.7,0)--(5.3,0);
        %\draw[thick] (0,-0.7)--(0,-1.3);
        \end{tikzpicture}}$
            \caption{Dynkin diagram of $D_4$, with the resolved nodes highlighted in black.}
    \label{fig:dynkinD4}
\end{figure}
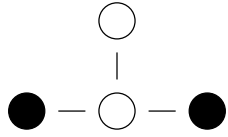

The analysis of non-isolated fourfold singularities constructed from base-changes of cDV singularities of type $D$ is precisely analogous to the $A$ series. The 3d $\mathcal{N}=2$ theory geometrically engineered by M-theory on them will generally give rise to flavor groups which are products of $U(1)$'s, as well as charged and uncharged chiral multiplets.\\
\indent For concreteness, let us show an example drawn from the $D$ cases:
\begin{equation}\label{D sing}
    x^2+zy^2 +z^{2n-1} -w^2t^2 = 0 \quad \subset \quad \mathbb{C}^5,
\end{equation}
which is singular at
\begin{equation}\label{sing lines}
    x = y = z = w = 0,\quad\quad x = y = z = t = 0.
\end{equation}
Far from the origin, the fourfold locally looks like the product of $\mathbb{C}$ with a cDV singularity, while on top the origin the singularity enhances.\\
\indent According to \eqref{D brane locus}, the corresponding brane locus reads:
\begin{equation}\label{D non iso locus}
    \Delta(\xi^2,w,t) = \xi^2(\xi^{2n-1}+wt)(\xi^{2n-1}-wt).
\end{equation}
We can easily construct the tachyon that encodes \eqref{D non iso locus} exploiting the toolbox from the previous sections. E.g.\ in the $n=2$ case it reads:
\begin{equation}\label{D tachyon}
    T = \left(
\begin{array}{cccc|cccc}
 \xi & 0 & 0 & 0 & 0 & 0 & 0 & 0 \\
 0 &  \xi & 1 & 0 & 0 & 0 & 0 & 0 \\
 0 & 0 &  \xi & w & 0 & 0 & 0 & 0 \\
 0 & t & 0 &  \xi & 0 & 0 & 0 & 0 \\
 \hline
 0 & 0 & 0 & 0 &  \xi & 0 & -t & 0 \\
 0 & 0 & 0 & 0 & -1 &  \xi & 0 & 0 \\
 0 & 0 & 0 & 0 & 0 & -w &  \xi & 0 \\
 0 & 0 & 0 & 0 & 0 & 0 & 0 &  \xi \\
\end{array}
\right),
\end{equation}
where we have highlighted its block structure, according to the conventions outlined in footnote \ref{bilinear form}. Tachyons for  $n$ higher than 2 are constructed in an identical fashion. It is easy to check that \eqref{D sing} admits a small resolution inflating two $\mathbb{P}^1$'s all along the singular lines \eqref{sing lines} as well as the origin, arranged as the black nodes in Figure \ref{fig:dynkinD4}.
A non-resolvable isolated singularity of the form
\begin{equation}\label{leftover sing}
    x_1^2+x_2^2+x_3^3+x_4^2+x_5^2 = 0,
\end{equation}
is leftover on top of the central fiber. This can be detected noticing that the part of the tachyon \eqref{D tachyon} that is not a multiple of the identity resides in a $A_2$ subalgebra of $D_4$ (corresponding to the white nodes in Figure \ref{fig:dynkinD4}), and computing its determinant when restricted to such subalgebra. We delve into the details of this construction in Appendix \ref{app:resolutions}.\\
\indent As a consequence, \eqref{D tachyon} preserves a $U(1)^2$ group, that plays the role of flavor group in the corresponding 3d $\mathcal{N}=2$ theory lying on top of the origin of the CY4 $x=y=z=w=t=0$. The physical content of such theory can be identified computing fluctuations of \eqref{D tachyon}, discarding the modes that are not localized at the origin. Carrying it out explicitly yields $\text{ceiling}\left(\frac{2n-1}{2}\right)$ uncharged modes and no charged modes.
Therefore:
\begin{figure}[H]
    \centering
             \centering
    \scalebox{0.9}{\begin{tikzpicture}
    \node at (0,0) {M-theory on};
   \node at (0,-0.6) {$x^2+zy^2+z^{2n-1}-w^2t^2=0$};
    \node at (3.2,-0.3) {$\Longleftrightarrow$};
    \node at (8,0) {3d $\mathcal{N}=2$ theory with $G_{\textit{flavor}}=U(1)^2$ of};
    \node at  (8.2,-0.6) {\text{ceiling}$\left(\frac{2n-1}{2}\right)$ uncharged chiral multiplets};
  % \node at  (8.2,-1.2) {\text{and 2$\cdot$floor}$\left(\frac{k-1}{2}\right)$ charged chiral multiplets};
    %%%%%
   \end{tikzpicture}}\;
    \caption*{}
    \label{fig:summary D non iso class}
    \end{figure}
    \vspace{-1.3cm}
This outcome is somewhat uninteresting, as there are no point particles charged under the flavor symmetry. Namely, all the localized modes come from the non-resolvable singularity \eqref{leftover sing}.\\

\indent Nonetheless, there exist CY4's built as ccDV singularities of type $D$ that sport abelian flavor symmetries, as well as chiral multiplets charged under it. One such example is:
\begin{equation}
  x^2+zy^2 + \left(z^3-t^2 w^2\right) \left(z^3-4 t^2 w^2\right) =0 \quad \subset \quad \mathbb{C}^5.
\end{equation}
Writing down the tachyon and computing its stabilizer and fluctuations proves the above-mentioned claim, yielding a $U(1)^2$ flavor group and 4 chiral multiplets charged under it (along with 4 uncharged chirals).

\subsection{Exceptional cases}\label{sec: E series}
\begin{figure}[H]
    \centering
$\scalebox{0.6}{\begin{tikzpicture}
        \draw[thick] (0,0) circle (0.4);
        %\node at (0,0) {\Large$\boldsymbol{\alpha_3}$};
        \draw[thick] (0.7,0)--(1.3,0);
        \draw[thick] (2,0) circle (0.4);
        %\node at (2,0) {\Large$\boldsymbol{\alpha_4}$};
        \draw[thick] (-0.7,0)--(-1.3,0);
        \draw[thick] (-2,0) circle (0.4);
        %\node at (-2,0) {\Large$\boldsymbol{\alpha_2}$};
        \draw[thick] (0,0.7)--(0,1.3);
        \draw[thick] (0,2) circle (0.4);
        \draw[thick, fill=black] (4,0) circle (0.4);
        \draw[thick, fill = black] (-4,0) circle (0.4);
        %\node at (0,2) {\Large$\boldsymbol{\alpha_6}$};
        \draw[thick] (2.7,0)--(3.45,0);
        %\draw[thick] (3.5,-0.5)--(4.5,-0.5)--(4.5,0.5)--(3.5,0.5)--cycle ;
      %  \node at (4,0) {\Large$\boldsymbol{\alpha_5}$};
        \draw[thick] (-2.7,0)--(-3.45,0);
        %\draw[thick] (-3.5,-0.5)--(-4.5,-0.5)--(-4.5,0.5)--(-3.5,0.5)--cycle ;
        %\node at (-4,0) {\Large$\boldsymbol{\alpha_1}$};
        %\draw[thick] (0,2.7)--(0,3.45);
        %\draw[thick] (-4.7,0)--(-5.3,0);
        %\draw[thick] (4.7,0)--(5.3,0);
        %\draw[thick] (0,-0.7)--(0,-1.3);
        \end{tikzpicture}}$
            \caption{Dynkin diagram of $E_6$, with the resolved nodes highlighted in black.}
    \label{fig:dynkinE6res}
\end{figure}
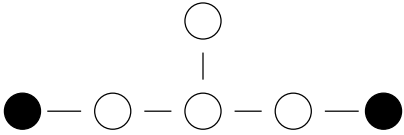

In order to explore M-theory geometric engineering on ccDV Calabi-Yau fourfolds with non-compact lines of \textit{exceptional} singularities intersecting (or enhancing) at a point, one can retrace the path adopted in the previous sections. As in Section~\ref{eq:isolExcepCases}, for the exceptional cases we need to use the adjoint scalar field $\Phi$ picture in order to derive the flavor group and the localized modes.\\
\indent Let us show an explicit example, given by a fourfold ccDV singularity built as a $E_6$ Du Val singularity deformed by two-complex parameters:
\begin{equation}\label{E6 non iso}
    x^2+y^3+z^4+w^3t^3 = 0 \quad \subset \quad \mathbb{C}^5.
\end{equation}
The fourfold \eqref{E6 non iso} has two non-compact lines of singularities at $x=y=z=w=0$ and $x=y=z=t=0$, intersecting on the origin of $\mathbb{C}^5$. An explicit adjoint scalar field whose Casimirs reproduce \eqref{E6 non iso} can be written as:
\begin{equation}\label{casimirs E6}
\begin{aligned}
    \Phi(w,t) \,=\,\,\,& e_{\alpha_6}+ite_{\alpha_2}-ie_{\alpha_3+\alpha_4}-te_{\alpha_2+\alpha_3}+\sqrt{3} we_{-\alpha_2-\alpha_3-\alpha_4-\alpha_6} \\ 
    &+\frac{1}{2} \left(-3 w+i \sqrt{3} w\right)e_{-\alpha_2-2\alpha_3-\alpha_4-\alpha_6}.
\end{aligned}
\end{equation}
Computing its fluctuations and stabilizer, one finds that \eqref{E6 non iso} admits a small resolution with two exceptional $\mathbb{P}^1$'s in the fiber of each singular point, arranged as in Figure \ref{fig:dynkinE6res}.
An additional non-resolvable isolated singularity of type $D$ from Table \ref{ccDV table}  is leftover on the fiber above the origin $x=y=z=w=t=0$. It is easy to check that this is the case noticing that \eqref{casimirs E6} resides in a $D_4$ subalgebra of $E_6$ (corresponding to the white nodes of Figure \ref{fig:dynkinE6res}), and analyzing the Casimirs of $\Phi$ restricted to that subalgebra. Physically, this leftover singular point produces the uncharged modes. The resulting effective 3d theory reads:
\begin{figure}[H]
    \centering
             \centering
    \scalebox{0.9}{\begin{tikzpicture}
    \node at (0,0) {M-theory on};
    \node at (0,-0.6) {$ x^2+y^3+z^4+w^3t^3 = 0$};
    \node at (3.2,-0.3) {$\Longleftrightarrow$};
    \node[align=left] at (8,-0.3) {3d $\mathcal{N}=2$ theory with $G_{\textit{flavor}}=U(1)^2$ of \\2 charged chiral multiplets and\\ 2 uncharged chiral multiplets};
    %%%%%
    \end{tikzpicture}}\;
    \caption*{}
    \label{fig:summary E non iso class}
    \end{figure}
    \vspace{-1.3cm}
Naturally, the same approach can be adopted to tackle the 3d theory arising from M-theory on ccDV singularities obtained as deformed $E_7$ and $E_8$ Du Val singularities.\\
\indent It is clear at this point of the discussion that further work is needed in order to fully classify non-isolated singularities giving rise to at most 2-cycles, and their corresponding 3d $\mathcal{N}=2$ theories.

\section{Conclusions}\label{sec: conclusion}
In this work we have started laying the foundations for a rigorous geometric engineering dictionary relating M-theory on non-compact CY4 with 3d $\mathcal{N}=2$ supersymmetric theories, reviewing the state of the art of the literature on terminal CY4, with both isolated and non-isolated singularities. We have initiated the exploration of the physical aspects of the dictionary from the simplest building blocks, namely terminal CY4 singularities admitting crepant resolutions with at most exceptional compact 2-cycles. These do not engineer 3d gauge theories, and yet are endowed with a non-trivial physical content, consisting of chiral multiplets that in some cases are charged under abelian flavor groups. We have carried out their investigation for CY4 constructed as deformed Du Val singularities of type $A$, $D$,~and~$E_6$. The absence of scales in the singular geometry and the presence of a suitable quasi-homogeneous action strongly hints at the fact that these theories flow to a SCFT in the deep IR.\\
\indent From the mathematical standpoint, we have presented a method to systematically compute the crepant resolutions of the classes of ccDV non-isolated singularities that have been introduced in this work, eschewing the need for an explicit blowup. Thanks to the tachyon condensation formalism, our techniques produce small partial resolutions of the ccDV fourfolds and clearly detect any leftover non-resolvable isolated singularities. We provide full details of our resolution recipe in Appendix \ref{app:resolutions}.\\

\indent Natural avenues for future work consist in generalizing our analysis for all the CY4 singularities with at most exceptional compact 2-cycles, providing a systematic classification, both from the mathematical point of view (as no exhaustive list has yet been provided), as well as from the 3d $\mathcal{N}=2$ physical perspective. It is expected that these theories shall contain at most chiral multiplets given by wrapped M2-branes, and no vector multiplets. In analogy with the case of rank-0 5d SCFTs, \textit{discrete} flavor symmetry groups could appear in the 3d theories engineered by ccDV singularities of type $D$ and $E$. Furthermore, the interplay between 0-form symmetries and 1-form symmetries, corresponding to unscreened defects encoded by M2-branes wrapping non-compact cycles, mandates further investigation. Striking another note related to symmetries, it would be worthy to investigate the realization of non-abelian symmetries in the ccDV fourfold context.\footnote{One such example is:
$
    x_1x_2 = x_3^nx_4^nx_5^n \,\, \subset \mathbb{C}^5
$,
that displays three non-compact surfaces supporting $A_{n-1}$ singularities. Blowing up these singularities gives rise to non-compact divisors that encode a flavor symmetry of type $A_{n-1}^{\oplus 3}$. We leave the analysis of this and related classes for future work.}
Similarly, it would be fairly straightforward to capture the consequences of the presence of T-branes \cite{Cecotti:2010bp} in our setups: they would generally produce a reduced spectrum of chiral multiplets, possibly breaking part of the flavor group. T-brane phases of our 3d $\mathcal{N}=2$ theories can be constructed with suitable tachyons according to the techniques introduced in the main text, analogously to what was done in the threefold cDV case in \cite{Collinucci:2021ofd,DeMarco:2021try,DeMarco:2022dgh}.\\
\indent From the point of view of the 3d $\mathcal{N}=2$ theories we have investigated, it is natural to analyze their moduli spaces. The chiral multiplets engineered by M2-branes wrapping $\mathbb{P}^1$'s in our resolved geometries become massless once the singular phase is reached and the volume of said $\mathbb{P}^1$'s shrinks to zero. This is a strong indication that the localized modes are genuine 3d modes with a finite kinetic term. As we have seen in Section \ref{sec: deformations}, giving a vev to such massless chirals deforms the singular geometry, and would then be interpreted as probing a branch of the moduli space of the 3d theory. It is well-known, though, that some of the isolated ccDV singularities examined in this work admit no normalizable deformations \cite{Gukov:1999ya}, and hence the fluctuations around non-trivial vev's are non-dynamical modes, but instead correspond to couplings of the theory. It is less clear what the deformation theory of the non-isolated cases is (it is crucial to remark that these are the only ones admitting a small resolution). This is a puzzle still awaiting a satisfactory solution, and that is in tune with recent progress on the physics of 5d $\mathcal{N}=1$ theories \cite{Acharya:2024bnt}. %, where it has been argued on mathematical grounds that no normalizable deformation exists.  
It would be interesting to explore this direction further both in the threefold and in the fourfold case.
Furthermore, superpotential couplings are less constrained in 3d $\mathcal{N}=2$ theories, with respect to the more supersymmetric 5d case; this may in principle lift part of the zero modes we found, making their number match the normalizable complex structure deformations. This question deserves future investigation.
 
%{\color{purple} RV: In realtà il puzzle \`e pi\`u rilevante: dal tachione o dal $\Phi$ si vede dove stanno i modi corrispondenti agli stati massless, se diamo vev a questi si vede che il CY4 viene deformato, quindi il legame tra massless modes e deformazioni risulta molto stretto; noi potremmo qui semplicemente notare che mentre alla singolarita ci aspettiamo che questi modi abbiano un termine cinetico finito, quando deformiamo GVW ci dicono che il termine cinetico esplode e i modi diventano non pi\`u dinamici. Il vev a sti modi lo puoi dare anche se non sono dinamici, semplicemente li interpreti come coupling, quindi non mi aspetto potenziali che liftino lo spazio dei moduli.}

\indent On a more mathematically-oriented note, the relationship of our 3d $\mathcal{N}=2$ theories with the enumerative invariants of CY4 is yet to be explored. Indeed, the physical content of these theories is given by BPS saturated states, that are in turn expected to be connected to GV, DT and PT invariants by a chain of dualities, although at present no clear theorem has been proven in this direction. In an increasing degree of difficulty, the next step for constructing the M-theory on CY4/3d $\mathcal{N}=2$ theory dictionary lies in the analysis of terminal CY4 with crepant resolution inflating compact 4-cycles. The corresponding theories need to be specified by additional non-geometrical data such as the background $G_4$ flux. This is the subject of current and upcoming work.%\footnote{In the compact case, it has been shown that the transition from a resolved to a deformed phase makes some $G_4$ flux emerge in the latter phase. This flux may in principle generate a superpotential, lifting some of the complex structure deformations.}
\\
\indent Finally, we have studied several examples of non-resolvable fourfold singularities. Non-resolvable codimension-3 singularities have proven to be particularly intriguing in F-theory model building \cite{Grimm:2010ez,Grimm:2011tb,Braun:2014nva,Arras:2016evy}.
It would be interesting to embed the non-resolvable codimension-4 singularities analyzed in this paper into compact Calabi-Yau fourfolds, and even more so into compact elliptically fibered Calabi-Yau fourfolds. These singularities naturally give rise to chiral multiplets without the need to turn on $G_4$ fluxes, offering greater flexibility in model building.

\section*{Acknowledgements}
We would like to thank Andreas P. Braun, Andr\'es Collinucci, Mario De Marco, Michele Graffeo, Michele Del Zotto, Sergej Monavari and Shani Nadir Meynet for useful discussions. The authors also thank Bobby Acharya, Sergio Benvenuti and Francesco Benini for fruitful discussions on related topics. R.V. acknowledges support by INFN Iniziativa Specifica ST\&FI. The work of A.S. is supported by the VR Centre for Geometry and Physics (VR grant No. 2022-06593). The work of A.S. is also co-funded by the European Research Council (ERC) under the European Union’s Horizon 2020 research and innovation program (grant agreement No. 851931).

\appendix

%\section{From GVW}

%\begin{itemize}
%    \item U(1) symmetry related to homogeneity, all isolated ccDV sing have a conic metric by Tian and Yau thm; relation to the fourfold singularity being canonical.
%    \item Fourfold isolated singularity with no vanishing cohomologies are the $A_n$ ccDV (called ADE sing in GVW). The other ccDV have vanishing cohomologies (that by GVW implies dynamical c.s. modes?).
%    \item For cases with resolutions, going from a resolved phase to a deformed phase, a G4-flux arises (to compensate the change in the Euler characteristic). So our cases with zero G4-flux have non-zero G4-flux along the compact 4-spheres blown up in the deformation. This flux may in principle lift some of the complex structure deformation (maybe the reason why we see less modes than c.s. deformations? but GVW page 15 say that there is no c.s. stabilisation for $A_n$ ccDV).
%\end{itemize}

\section{Crepant resolutions of ccDV singularities}
\label{app:resolutions}

In this Appendix we schematically summarize our strategy for rigorously computing the crepant resolutions of the ccDV singularities examined in this work, thanks to the physics-inspired tachyon condensation formalism.\\
\indent We wish to resolve a generic ccDV fourfold, that might admit isolated or non-isolated singularities. Such singular fourfold can always be written as:
\begin{equation}\label{X sing}
    X: \quad P_{\mathfrak{g}}(x,y,z) +h(x,y,z,w,t) = 0,
\end{equation}
where $P_{\mathfrak{g}}$ is a Du Val singularity labelled by $\mathfrak{g} \in ADE$ and $h$ is a function such that $h(x,y,z,0,0) =0$. In this work, we have shown how to construct an adjoint scalar field $\Phi$ that encodes a large subclass of the singularities of type \eqref{X sing}, via the spectral equations in Table \ref{table spectral} and \ref{E table spectral}. From the technical point of view, no hurdle prevents the construction of a tachyon for \textit{all} ccDV singularities. Thus, there is a one-to-one correspondence between $\Phi$ and a ccDV singularity\footnote{Here we are consciously forgetting about possible T-brane phases, that would correspond to a different $\Phi$.}.\\
\indent We claim that the crepant resolution (if any) of a ccDV singularity $X$ can be completely characterized by the scalar field $\Phi$. In general, a ccDV singularity can admit at most a small resolution, as proven thanks to the theory of Grothendieck-Springer resolutions and briefly summarized at the end of Section~\ref{sec: non iso curve singularities}. After this small resolution, there can be leftover non-resolvable singularities. Let us briefly outline the main possible cases:
\begin{itemize}
    \item $X$ is an isolated singularity. No crepant small resolution exists, due to \hyperlink{thm4}{Theorem 4}.
    \item $X$ is a non-isolated singularity constructed as the intersection of two singular lines. Suppose that the fiber along the non-compact singular lines are \textit{resolvable} cDV singularities. Then $X$ admits a small resolution, possibly with leftover isolated singular points of the type appearing in Table \ref{ccDV table} on the central fiber.
    \item $X$ is a non-isolated singularity constructed as the intersection of two singular lines. Suppose that the fibers along the non-compact singular lines are \textit{non-resolvable} cDV singularities. Then $X$ does not admit a small resolution.
    \item If $X$ is a non-isolated singularity constructed as the intersection of more than two singular lines, a generalization of the previous criteria applies.
\end{itemize}
We claim that the scalar field $\Phi$ associated to $X$ \textit{automatically} encodes its small resolution, if it exists, and further identifies any leftover isolated singular points in the partially resolved phase. Let us see how this concretely comes about.\\

\indent As we have recalled in the main text, a ccDV singularity is labelled by an ADE algebra $\mathfrak{g}$, with simple roots $\alpha_i$. Define also a basis for the Cartan generators of $\mathfrak{g}$ employing the duals of the simple roots:
\begin{equation}\label{cartan basis}
    \mathfrak{t} = \langle \alpha^*_1,\ldots,\alpha^*_r \rangle,
\end{equation}
with $r$ the rank of $\mathfrak{g}$. The scalar field $\Phi$ is represented as a matrix transforming in the adjoint representation of $\mathfrak{g}$. Here, we are considering $\Phi$ as the scalar field that produces the maximal spectrum and the largest flavor symmetry in 3d\footnote{Finding such $\Phi$ is the only step in the algorithm that requires a creative input.}.\\
\indent In general, $\Phi$ will lie in a subalgebra $\mathfrak{h}$ of $\mathfrak{g}$:
\begin{equation}
    \Phi \in \mathfrak{h} \subseteq \mathfrak{g}, \quad \text{ with } \mathfrak{g} \in ADE.
\end{equation}
The centerpiece of our construction claims that \textit{the commutant of $\Phi$ dictates the small resolution of the corresponding ccDV singularity}. As we are examining ccDV singularities with at most singular lines, the commutant $\mathfrak{m}$ of $\Phi$ is always abelian, and can be expanded in the basis \eqref{cartan basis} as:
\begin{equation}\label{commutant}
    \mathfrak{m} = \langle \alpha^*_{i_1} ,\ldots,\alpha^*_{i_k}\rangle,
\end{equation}
for some integer $k$. In geometrical terms, \eqref{commutant} is telling us that each of the roots labelled by $i_1,\ldots,i_k$ will give rise to a $\mathbb{P}^1$ in the partially resolved phase $\tilde{X}$. Thus, a crucial requirement for a ccDV singularity to possess a small resolution is that the scalar field $\Phi$ that describes it lies in a subalgebra of $\mathfrak{g}$ with a non-trivial commutant. We have shown three explicit examples of this technique in the main text, in Section~\ref{sec: A series v2}, \ref{sec: D series} and \ref{sec: E series}, respectively for the $A$, $D$ and $E_6$ cases.\\
\indent In order to detect any leftover non-resolvable isolated singularity that sticks around after the small crepant partial resolution, namely in the $\tilde{X}$ phase, we have to examine the structure of $\mathfrak{h}$ in more detail. In all generality, it is the sum of simple addends:
\begin{equation}
    \mathfrak{h} = \bigoplus_i \mathfrak{h}_i \oplus \mathfrak{m}.
\end{equation}
Each of the simple addends $\mathfrak{h}_i$ corresponds to a subalgebra of $\mathfrak{g}$, that can be readily identified from the Dynkin diagram. In the main text we have shown that the singular fourfold equation is recoverable from $\Phi$ thanks to the spectral equations in Table \ref{table spectral} and \ref{E table spectral}, whose only non-trivial inputs are \textit{the Casimirs of $\Phi$}. In order to identify the leftover singularities after the small resolution, it suffices to compute the Casimirs of $\Phi$ \textit{restricted to each of the simple subalgebras $\mathfrak{h}_i$}. Such Casimirs will uniquely charaterize, through the spectral equation, a fourfold of ccDV type labelled by the ADE algebra $\mathfrak{h}_i$, that might or might not be singular. If it is singular, it corresponds to one of the singularities appearing in Table \ref{ccDV table}.\\
\indent We can finally recap the resolution procedure of a ccDV singularity as follows:
\begin{equation*}
\begin{array}{c}
   \text{Singular fourfold }X \quad \longleftrightarrow \quad \Phi \in \bigoplus_i \mathfrak{h}_i \oplus \mathfrak{m} \subseteq \mathfrak{g}, \quad \begin{array}{l} 
    \mathfrak{g} \in ADE\\
    \mathfrak{h}_i \text{ simple algebras}\\ \mathfrak{m} \text{ abelian}\\
    \end{array}   
    \vspace{0.3cm}
\\ 
   \hline      
    \mathfrak{m} \quad \longleftrightarrow \quad \text{subset of simple roots } \{\alpha_{i_1} ,\ldots,\alpha_{i_k}\} \quad \longleftrightarrow \quad \\
    \longleftrightarrow \quad \text{$\mathbb{P}^1$'s in $\tilde{X}$, arranged like the roots $\{\alpha_{i_1} ,\ldots,\alpha_{i_k}\}$ in the $\mathfrak{g}$ Dynkin diagram} 
    \vspace{0.3cm}
    \\
   \hline
    \text{Casimirs of $\Phi$ restricted to $\mathfrak{h}_i$} \quad \longleftrightarrow \quad \text{leftover isolated singularities in $\tilde{X}$ from Table \ref{tab:isolated summary}}
    \Tstrut
\end{array}
\end{equation*}

\providecommand{\href}[2]{#2}

\end{document}